\documentclass[preprint]{aastex631}
\usepackage{float}
\usepackage{hyperref}
\usepackage{amsmath}
\usepackage{framed}

\renewcommand{\micron}{$\mu$m}
\newcommand{\ab}{$\sim$}
\renewcommand{\farcs}{.$^{\prime\prime}$}

\newcommand{\msun}{M$_{\odot}$}

\newcommand{\kms}{\mbox{km s$^{-1}$}}

\newcommand{\ceo}{C$^{18}$O}
\newcommand{\htco}{H$_{2}$CO}
\newcommand{\co}{$^{12}$CO}
\newcommand{\cso}{C$^{17}$O}

\newcommand{\tco}{$^{13}$CO}

\newcommand{\beam}{beam$^{-1}$}

\renewcommand{\arcsec}{$^{\prime\prime}$}
\renewcommand{\deg}{$^{\circ}$}

\renewcommand{\added}[1]{#1}
\renewcommand{\deleted}[1]{}
\renewcommand{\edit}[2]{#2}
\newcommand{\addedtwo}[1]{#1}
\newcommand{\deletedtwo}[1]{}

\received{March 1, 2023}
\revised{Dec 11, 2023}
\accepted{Dec 11, 2023}

\submitjournal{APJ}

\begin{document}

\title{The Disk Orientations of Perseus Protostellar Multiples at $\sim$8~au Resolution}

\author[0000-0002-9239-6422]{Nickalas K. Reynolds}
\affiliation{Homer L. Dodge Department of Physics and Astronomy, The University of Oklahoma, 440 W Brooks St, Norman, OK, 73019 USA}

\author[0000-0002-6195-0152]{John J. Tobin}
\author[0000-0002-9209-8708]{Patrick D. Sheehan}
\affiliation{National Radio Astronomy Observatory, 520 Edgemont Rd, Charlottesville, VA 22903 USA}

\author[0000-0001-7474-6874]{Sarah I. Sadavoy}
\affiliation{Department of Physics, Engineering Physics and Astronomy, Queen’s University, Kingston, ON, K7L 3N6, Canada}

\author[0000-0002-4540-6587]{Leslie W. Looney}
\affiliation{Department of Astronomy, University of Illinois, 1002 W. Green St., Urbana, IL 61801, USA }

\author[0000-0001-7474-6874]{Kaitlin M. Kratter}
\affiliation{Department of Astronomy, /Steward Observatory, University of Arizona 933 North Cherry Avenue, Rm. N204, Tucson, AZ 85721-0065}

\author{Zhi-Yun Li}
\affiliation{Department of Astronomy, University of Virginia, 530 McCormick Rd., Charlottesville, VA 22903, USA}

\author[0000-0003-3172-6763]{Dominique M. Segura-Cox}\thanks{NSF Astronomy and Astrophysics Postdoctoral Fellow}
\affiliation{Department of Astronomy, University of Texas, 2515 Speedway Stop Austin, TX 78712}
\affiliation{Max Planck Institute for extraterrestrial Physics,Giessenbachstrasse 1, D-85748 Garching, Bayern, Germany}

\author[0000-0001-5272-5888]{Nathan A. Kaib}
\affiliation{Homer L. Dodge Department of Physics and Astronomy, The University of Oklahoma, 440 W Brooks St, Norman, OK, 73019 USA}

\begin{abstract}

We present a statistical characterization of circumstellar disk orientations toward 12 protostellar multiple systems in the Perseus molecular cloud using the Atacama Large Millimeter/submillimeter Array at Band~6 (1.3~mm) with a resolution of \ab25~mas ( \ab8~au). This exquisite resolution enabled us to resolve the compact inner disk structures surrounding the components of each multiple system and to determine the projected 3-D orientation of the disks (position angle and inclination) to high precision. We performed a statistical analysis on the relative alignment of disk pairs to determine whether the disks are preferentially aligned or randomly distributed. \added{We considered three subsamples of the observations selected by the companion separations, $a<$100~au, $a>$500~au, and $a<$~10,000~au. We found for the compact ($<100$~au) subsample, the distribution of orientation angles is best described by an underlying distribution of preferentially aligned sources (within 30\deg) but does not rule out distributions with 40\%\space misaligned sources. The wide companion ($>$500~au) subsample appears to be consistent with a distribution of 40\%-80\%\space preferentially aligned sources. Similarly, the full sample of systems with companions (a$<10,000$~au) is most consistent with a fractional ratio of at most 80\%\space preferentially aligned source and rules out purely randomly aligned distributions.  Thus our results imply the compact sources ($<$100~au) and the wide companions ($>$500~au) are statistically different.}

\end{abstract}


\section{Introduction}\label{sec:introduction}
Recent studies in the past several decades have shown nearly half of all solar-type star systems are multiples \citep{raghavanmcalister2010, 2013duchenekraus, 2017moestefano, 2022offnermoekrattertobin}. It has been discovered that stellar multiplicity is even more common for young stars \citep{1994mathieu, 2013chenarce, 2020tobinsheehanmegeath}, and protostars in the midst of the stellar assembly process have the highest multiplicity fractions \citep{2008connelleyreipurth,2013chenarce,2016tobinlooney}. During the earliest stages of star formation, the deeply-embedded protostellar phase, the largest reservoir of mass is available to form multiples \citep{2002tohline}. This is the stage of stellar evolution that must be examined to reveal the origins of stellar multiplicity.

Multiple star systems are thought to primarily form via two processes that operate on distinct scales: massive disks undergoing disk fragmentation on 10s-100s~au scales \citep[e.g., ][]{2010krattermatzner} and turbulent core fragmentation on $\sim$1000s~au scales \citep[e.g., ][]{2010offnerkratter}. These processes can operate simultaneously, possibly giving rise to populations of close ($<$500~au) and wide ($>$1000~au) multiple systems \citep{2016tobinlooney}. However, while the scales by which these processes form multiples are distinct, the systems formed via turbulent fragmentation may migrate to $\sim$100s~au separations (or less) in $\sim$10s of kyr \citep{2010offnerkratter, 2019leeoffner} depending on their relative gravitational attraction with the core and the relative velocities of the sources at the times of formation. This makes it difficult to uniquely identify the dominant formation mechanism from separation measurements alone. Studies of protostellar multiplicity within the Orion molecular cloud by \citet{2022tobinoffnerkratter} found that current simulations of turbulent collapse alone did not account for all of the observed multiples found between 20-500~au, and thus an additional mechanism was needed to explain the observations. Meanwhile, \citet[][]{2016murillodishoeck} characterized the relative evolutionary states for wide and compact YSO multiples using SED modeling and found \ab33\%\space of multiple systems were inconsistent with ``co-eval'' formation mechanisms.

Distinguishing if there is a primary mechanism for close multiple star formation is important for understanding the origin of stellar multiples, their evolution, and the potential impact they might have on their circumstellar disks and the planet formation potential. A close multiple system formed within a circumbinary disk may undergo relatively smooth evolution, while a close multiple system that forms as a result of migration from large separations can greatly disturb the system, leading to misalignment of outflows \citep{2016offnerdunham} and possibly disrupting accretion disks.

The VLA Nascent Disk And Multiplicity (VANDAM) Survey characterized the multiplicity of the entire protostar population in the Perseus molecular cloud \citep{2016tobinkratter}, finding 17 multiple systems with separations less than 2\farcs0~(600~au) out of 90 sources observed. This sample of close multiples was followed-up with ALMA observations of 1.3~mm continuum and molecular lines that likely trace disk kinematics (\tco, \ceo, SO, and \htco), enabling the most likely formation mechanism to be inferred for 12 systems \citep{2018tobinlooneyli}. Eight out of 12 systems were found to be consistent with disk fragmentation and four were inconsistent. The systems that were consistent with disk fragmentation had apparent rotating circumbinary structures surrounding the binary/multiple system. However, even with this evidence, turbulent fragmentation could not be completely ruled-out. This is because a system formed from turbulent fragmentation could migrate inward, interact, and form a close multiple system with a new circumbinary disk \citep[e.g.,][]{2018bate}. 

However, the compact circumstellar disks around each component of the close multiple systems could provide more definitive evidence on the formation mechanism. If the close multiple system is formed via disk fragmentation, the angular momentum axis of each component and the circumbinary disks should be relatively aligned due to forming within a common disk with the same net angular momentum \citep{2016offnerdunham, 2018bate}. On the other hand, \citet{2019lee}\space and  \citet{2022offnertaylormarkeychen}\space found that in simulations of turbulent core collapse, multiples preferentially formed as randomly aligned systems whose (mis-)alignment angles would persist throughout the calculation, suggesting randomly distributed alignment angles should be a signature for this formation mechanism.

\added{These primordial alignments can further evolve via dynamical interactions to either mis-align or re-align at later stages. Highly inclined alignments with respect to the orbital plane can quickly decay, on order of 10$^4$-10$^5$~years, while more moderate misalignments decay over 10$^6$-10$^7$~years \citep[typical of 1~\msun\space companions on \ab100~au separations, e.g., ][]{2000batebonnellclarke}. Further studies of the more evolved protoplanetary disks are consistent with a greater likelihood of misaligned companions but these system probably experienced many orbital timescales, thus undergoing dynamical evolution \citep{2004jensenmathieudonar,2014jensenakeson, 2022rotamanaramiotello}. Furthermore, \citet{1996larwoodnelson}\space found the disk inclination angle and orbital angular momentum axis evolve on timescales of order the viscous timescale, thus it is not clear if the observed relative alignment angles are inherited from the formation or if the angles are a product of the dynamical evolution. Making surveys of the circumstellar disks at such young evolutionary stages  critical for breaking the degeneracy between the primordial alignments versus the more evolved protoplanetary disks.}

At such a young age, protostars are deeply embedded, making direct measurements of the stellar rotation axis impossible. \citet{2010batelodatopringle} found that the stellar rotation axis (the inferred stellar angular momentum axis) would not differ significantly from the inner disk rotation axis ($<$5\deg). \added{The outflow position angle can be a proxy for the angular momentum axis, can be difficult to separate for compact systems. These outflows may be entangled and/or misaligned from the rotation axis due to n-body interactions \citep[][]{2022ohashi}.}\deleted{While the other method commonly used for determining the angular momentum axis, outflow position angles, can be difficult for compact systems, whose outflows may be entangled or be misaligned with the rotation axis \citep[][]{2022ohashi}; outflows from the individual components were generally not obvious in the observations from \citet{2018tobinlooneyli}.} \added{\citet{2018tobinlooneyli} confirmed and resolved the companion separations, but given the resolution were not able to resolve the compact circumstellar disks or the individual outflows around the sources. Furthermore, \citet{2018seguracoxlooney} conducted high resolution VLA observations and identified Per-emb-5 potential compact multiple, requiring high-sensitivity follow-up to confirm the multiplicity.} Thus high-resolution and high-sensitivity studies of compact and wide multiples are needed to accurately determine the angular momentum vectors of the disks.


We carried out a novel method of empirically testing multiple protostar formation mechanisms by observing \added{12} known wide and compact multiples with 25~mas angular resolution (\ab8~au), with the ability to resolve small protostellar disks. This type of survey can best recover the projected disk rotation axis, the implied orientation vector, of the nascent circumstellar disk. \added{Similar studies have been conducted in the past, observing the polarization angle of the sources to infer the circumstellar disk alignments but are highly sensitive to the amount of intervening interstellar polarization alignments which may contaminate the resulting distribution of angles \citet{2004jensenmathieudonar}.} 

We present our findings of 12 protostellar multiple systems within the Perseus molecular cloud and detail the collection of observations in Section~\ref{sec:observations}. We present our analysis of the protostellar sample in Section~\ref{sec:statanalysis}. Further, we interpret our findings in the broader aspects of star formation and with the specific sources of the sample in Section~\ref{sec:discussion}.

\subsection*{Definitions}
For consistency we utilized the following definitions for this work:
\begin{enumerate}
    \item \textbf{source}: A single protostar that can have a compact and/or extended disk.
    \item \textbf{system}: A collection of sources within a defined separation that may be interacting.
    \item \textbf{aligned}: A pair of sources whose dot product of the orientation vectors would correspond to a value of $<$30\deg\space\citep{2019leeoffner}.
    \item \textbf{misaligned}: A pair of sources whose dot product of the normalized orientation vectors would correspond to a value of $>$30\deg.
\end{enumerate}
\added{The definitions of alignment are for ease of qualitative referencing and is not relied on for the analysis detailed in Section~\ref{sec:statanalysis}. In the analysis, we utilized the alignment angles and the corresponding observational errors. The demarcation of alignment between two orientation vectors of 30\deg\space are chosen to remain consistent with studies of simulated data \citep{2019leeoffner}.}

\section{Observations and Data Analysis} \label{sec:observations}
The Atacama Large Millimeter/submillimeter Array (ALMA) is a state-of-the-art interferometer located on Llano de Chajnantor in the Atacama region of Chile at an elevation of \ab5000~meters). We conducted observations of protostellar multiple systems in Perseus primarily using Band 6 (1.3~mm) with some supplementary observations in Band 3 (3~mm).

\subsection{Band 6~(1.3~mm) Observations}
The observations were taken as part of project 2019.1.01425.S  with 48 antennas included between 2019 September 11-18 at Band 6~(1.3~mm) toward 12 protostellar systems in the Perseus molecular cloud (d\ab300~pc). The observations were carried out in the most extended configuration C-9/10 (baselines 150~m\ab14.9~km) and have an effective angular resolution of 42~mas$\times$23~mas\space to 66~mas$\times$29~mas, with a continuum sensitivity of 13-76~$\mu$Jy~\beam, when reconstructed with the Briggs robust 1 weighted imaging.  The correlator was configured with 3 spectral windows setup for 1.875~GHz bandwidth and 3840 channels and four spectral windows used 117.19~MHz bandwidth and centered on the $^{13}$CO ($J=2\rightarrow1$), C$^{18}$O ($J=2\rightarrow1$), SO ($J=6(5)\rightarrow5(4)$), and SiO ($J=5\rightarrow4$) transitions. However, the spectral lines were not well-detected given that the integration times were chosen for continuum sensitivity.

The first set of observations for 5 of the sources took place across a 1.5-hour block, with each time-on-source $\approx$9~minutes (scans\ab9.04~s). The remaining 7 sources were observed in 3 execution blocks (EBs), across 3 days, with the average time-on-source \ab14~minutes for the 3 EBs combined. While the absolute flux density scale is expected to be accurate to $\sim$10\%, for the purpose of the results presented, all flux uncertainties only consider statistical uncertainties. A summary of the scheduled observations and execution blocks is given in Tables~\ref{table:13mmpointing}~and~\ref{table:13mmsb}.

    \begin{deluxetable}{lrcccccccc}
    \tablecaption{1.3~mm Pointings}
    \tablehead{
\colhead{Name} & \colhead{Other names} & \colhead{$\alpha$} & \colhead{$\delta$} & \colhead{beam} & \colhead{RMS} & \colhead{S/N} & \colhead{Class} & \colhead{L$_{bol}$} & \colhead{T$_{bol}$}\\
\colhead{} & \colhead{} & \colhead{(J2000)} & \colhead{(J2000)} & \colhead{(mas$\times$mas)} & \colhead{($\mu$Jy~beam$^{-1}$)} & \colhead{} & \colhead{} & \colhead{(L$_{\odot}$)} & \colhead{(K)}
}
    \startdata
L1448 IRS1 & \nodata & 3:25:9.45 & 30:46:21.84 & 43$\times$23&13& 391& I & 2 & None\\
Per-emb-2 & IRAS 03292+3039 & 3:32:17.93 & 30:49:47.7 & 61$\times$27&38& 19& 0 & 0.9 & 27\\
Per-emb-5 & IRAS 03282+3035 & 3:31:20.94 & 30:45:30.19 & 61$\times$28&47& 101& 0 & 1.3 & 32\\
NGC1333 IRAS2A & Per-emb-27 & 3:28:55.57 & 31:14:36.92 & 62$\times$27&53& 368& 0/I & 19 & 69\\
Per-emb-17 & L1455 IRS1 & 3:27:39.11 & 30:13:2.98 & 42$\times$23&13& 429& 0 & 4.2 & 59\\
Per-emb-18+ & NGC1333 IRAS7 & 3:29:11.27 & 31:18:30.99 & 63$\times$28&17& 181& 0 & 2.8 & 59\\
Per-emb-21+ & \nodata & 3:29:10.67 & 31:18:20.09 & 63$\times$28&17& 359& 0 & 6.9 & 45\\
Per-emb-22 & L1448 IRS2 & 3:25:22.42 & 30:45:13.16 & 43$\times$23&13& 448& 0 & 3.6 & 43\\
L1448 IRS3B+ & Per-emb-33 & 3:25:36.32 & 30:45:14.81 & 43$\times$23&14& 169& 0 & 8.3 & 57\\
L1448 IRS3A+ & \nodata & 3:25:36.5 & 30:45:21.83 & 43$\times$23&14& 365& I & 9.2 & 47\\
L1448 IRS3C & L1448 NW & 3:25:35.68 & 30:45:34.26 & 43$\times$23&20& 192& 0 & 1.4 & 22\\
Per-emb-35 & NGC1333 IRAS1 & 3:28:37.1 & 31:13:30.72 & 62$\times$28&18& 307& I & 9.1 & 103\\
NGC1333 IRAS2B & Per-emb-36 & 3:28:57.38 & 31:14:15.67 & 66$\times$29&37& 294& I & 5.3 & 106\\
SVS13A+ & Per-emb-44 & 3:29:3.77 & 31:16:3.71 & 64$\times$28&76& 194& 0/I & 32.5 & 188\\
SVS13A2+ & \nodata & 3:29:3.39 & 31:16:1.53 & 64$\times$28&76& 59& 0/I & 32.5 & 188\\
SVS13B+ & \nodata & 3:29:3.08 & 31:15:51.64 & 64$\times$28&76& 32& 0 & 1 & 20\\
\enddata
    \tablecomments{References: \citet{2009enochevans, 2014sadavoydifrancesco}}
    \tablecomments{L1448IRS3B also contains L1448IRS3A within the field of view. SVS13A also contains SVS13B within the field of view. Per-emb-18 also contains Per-emb-21 within the field of view.}
    \tablecomments{RMS is specified as the root-mean-squared of the Briggs robust 1 weighted image with the upper 95\%\space of emission clipped. The beam specified is the self-calibrated, multi-frequency synthesis \textit{clean}\space beam using Briggs robust 1 weighting.  S/N is the signal-to-noise ratio defined as the emission peak divided by the respective RMS. Class, T$_{bol}$, and L$_{bol}$\space are given in \cite{2018tobinlooneyli, 2008connelley}.}\label{table:13mmpointing}

    \end{deluxetable}

\begin{deluxetable}{lccccc}
\tablecaption{1.3~mm Scheduling Block}
\tablehead{
\colhead{Identifier} & \colhead{\# E.B.} & \colhead{Date} & \colhead{Phase Cal.} & \colhead{Bandpass Cal.} & \colhead{Flux Cal.}\\
\colhead{} & \colhead{} & \colhead{(Sept 2021)} & \colhead{} & \colhead{} & \colhead{}
}
\startdata
(1) & 1 & 11 & J0336$+$3218 & J0237$+$2848 & J0341$+$3352\\
(2) & 3 & 13,16,18 & J0336$+$3218 & J0237$+$2848 & J0338$+$3106\\
\enddata
\tablecomments{Identifier (1) corresponds to the Member Observation Unit Set (MOUS) of uid://A001/X1465/Xd60 and (2) corresponds to uid://A001/X1465/Xd63}
\tablecomments{\# E.B. is the number of execution blocks.}\label{table:13mmsb}
\end{deluxetable}

The raw visibility data were calibrated by the North American ARC staff using Common Astronomy Software Applications (CASA) version 6.2.1 automated pipeline.  The high sensitivity of the observations (pre-self-calibration\ab40~$\mu$Jy) and signal-to-noise ratios (S/N) of at least 50, enabled self-calibration to be attempted for all sources except for Per-emb-2. We performed 4 to 5 rounds of phase-only self calibration, with the first round of intervals starting at the full length of the EB, then round two of intervals starting at the full length of the on-source scans, then progressing to 18.14~s, 9.07~s, and ending at the single integration timestep. Per-emb-18, Per-emb-35, NGC1333~IRAS2B, and L1448~IRS3B were unable to be phase self-calibrated down to the shortest timestep due to the S/N degrading, but were phase-only self-calibrated down to 9.07~s. The final average sensitivity resulting from the phase-only self calibrations was $\sim$30~$\mu$Jy\space and an average increase of S/N by a factor of 1.5. A summary of the observations is detailed in Table~\ref{table:13mmpointing}.

The data were imaged using CASA version 6.5.0-15 with the task \textit{tclean}, and the images using Briggs weighting with the robust parameter 1.0 are shown in Figures~\ref{fig:cont}~and~\ref{fig:contzoom} with image sizes of 9000$\times$9000\space and 4~mas~per~pixel. All images are shown with square-root stretch colormaps and a common RMS value of 20~$\mu$Jy~\beam. To restore the images, we used the Multi-(Taylor-)term Multi-Frequency Synthesis (MTMFS) with scale sizes of 0, 5, and 20~pixels and 2 Taylor terms. The scale sizes were chosen to recover dominant features in the disk and correspond to physical sizes of point source, typical size of the beam minor axis, and 2$\times$~the typical beam major axis. We utilized the ``auto-multithresh'' masking technique to non-interactively mask and clean the data\added{ in a reproducible manner} by defining the sidelobe threshold sub-parameter to be 2.0.\added{ The final images were checked by simultaneously cleaning the data with a conservative user-defined mask.}

\subsection{Band 3~(3~mm) Observations}
ALMA Band 3 data were taken toward three targets, a subset of the Band 6 observations: Per-emb-2, Per-emb-18, and Per-emb-5, within ALMA project 2016.1.00337.S. Data were taken in two ALMA configurations, C40-6 (2017 Aug 1) and C40-9 (2017 September 27), summarized in Table~\ref{table:3mmpointings}. For both datasets, the phase calibrator was J0336+3218, the flux calibrator was J0238+1636 and the bandpass calibrator was J0237+2848. The baselines sampled in the combined dataset ranged from 16~m to 14,500~m.

The correlator was configured with 3 spectral windows setup for 1.875 GHz bandwidth and 128 channels and two spectral windows used 58.59 MHz bandwidth and centered on the $^{13}$CO and C$^{18}$O ($J=1\rightarrow0$) transitions. However, the spectral lines were not well-detected given that the integration times were chosen for continuum sensitivity. The central frequency of the observations is $\sim$102 GHz.

The data were pipeline processed by the observatory using the pipeline included in CASA 4.7.2 (r39732). The data were also self-calibrated and imaged using this same version of CASA. The C40-6 data went through 3 rounds of phase-only self-calibration with solution intervals of one scan, 24.15s, and 6.05s (a single integration). The C40-9 data also went through three rounds of phase-only self-calibration but the second interval used a 12.10s solution interval, and the final solution interval of 6.05s corresponded to 3 integrations. 

The final self-calibrated data were imaged together using the CASA task \textit{clean} with image sizes of 2048$\times$2048 pixels and 5~mas pixels. We made use of MTMFS imaging given the wide fractional bandwidth, \added{restoring the images with super-uniform weighting scheme to closely match the beamsize of the 1.3~mm observations,} and \deleted{the images were cleaned }interactively cleaned using hand-drawn masks. The images were cleaned down to $\sim$1.5$\times$ the noise in each image (Figure~\ref{fig:3mmcont}).

\begin{deluxetable}{lrcccccccc}
\tablecaption{3~mm Pointings}
\tablehead{
\colhead{Name} & \colhead{$\alpha$} & \colhead{$\delta$} & \colhead{beam} & \colhead{RMS} & \colhead{S/N}\\
\colhead{} & \colhead{(J2000)} & \colhead{(J2000)} & \colhead{(mas$\times$mas)} & \colhead{($\mu$Jy~beam$^{-1}$)} & \colhead{}
}
\startdata
Per-emb-2 & 3:32:17.93 & 30:49:47.73 & 108$\times$48&18& 75\\
Per-emb-5 & 3:31:20.94 & 30:45:30.23 & 108$\times$48&18& 168\\
Per-emb-18+ & 3:29:11.26 & 31:18:31.04 & 90$\times$41&27& 34\\
Per-emb-21+ & 3:29:10.67 & 31:18:20.14 & 187$\times$110&10& 475\\
\enddata
\tablecomments{Per-emb-18 also contains Per-emb-21 within the field of view.}
\tablecomments{RMS is specified as the root-mean-squared of the superuniform weighted image with the upper 95\%\space of emission clipped. The beam specified is the self-calibrated, multi-frequency synthesis \textit{clean}\space beam using superuniform weighting. S/N is the signal-to-noise ratio defined as the emission peak divided by the respective RMS.}\label{table:3mmpointings}
\end{deluxetable}

\subsection{Gaussian Fitting the \textit{uv}-visibilities}
For the most compact companion sources (Per-emb-18, L1448~IRS3B, L1448~IRS3C, and SVS13A), the compact disk emission is only slightly larger than the size of the beam, so the deconvolved PA and \textit{i}\space derived from image-plane analysis will be less well-constrained. Moreover, we desired an alternate fitting method to measure the source parameters independent of the images produced with the CLEAN algorithm.\added{ For completeness, we conducted the image-plane analysis in Appendix~\ref{sec:imagefit} and find the results between the \textit{uv}-visibility and image plane analysis are consistent}\addedtwo{ within 3~$\sigma$}\added{.}

In order to utilize the full spatial constraints afforded by the observations, we constructed a number of Gaussians equal to the number of sources in the \textit{uv}-visibilities. Similar techniques were applied to protoplanetary disks \citep{2022jenningsdhsarp, 2022jenningstaurus} to recover substructure with $>2\times$\space longer effective baselines than the reconstruction from CLEAN by fitting the visibilities directly.  We constrained the sources using Bayesian inference, fitting multi-component 2-D Gaussians to the visibilities of each individual source using \citep[\textit{emcee}, \textit{dynesty}, \textit{pdspy}; ][]{emcee, 2020dynesty, 2022pdspy}. We restricted the \textit{uv}-visibilities fitting to scales smaller than 0\farcs5 (by restricting the \textit{uv}-distance $>$400~k$\lambda$) to ensure we fit the compact disk and not the extended emission of the envelope or circum-multiple material that was not resolved out in the observations. We also limited the fit phase center to be within 0\farcs5 of previously published results, and in the cases of new detections, we utilized the centering from the CASA task \textit{imfit} to form the \textit{prior}\added{ (a summary of the \textit{imfit} results and the comparison with the \textit{uv}-visibility results is described in Appendix~\ref{sec:imagefit})}.

A summary of the fitted parameters is provided in Table~\ref{table:13mmuvfits}\space, and a summary of the projected 3D orientation vectors solved from fitting the \textit{uv}-visibilities is shown in Figure~\ref{fig:jveczoom}. The errors reported are derived as the 1~$\sigma$\space uncertainty from the median of the sampled posterior.

While all sources could be described by a Gaussian, the source L1448 IRS3A was best described with a ring (see Appendix~\ref{sec:specificsources}). The detailed analysis of L1448 IRS3A falls outside of the scope of this paper and we leave analysis of its disk structure for a future paper.
While the \textit{uv}-visibilities enable a more complete picture of the system and accurate representation of the sources, without bias from the inherent beam geometries and subsequent clean procedure.

    \begin{longrotatetable}
    \begin{deluxetable}{lrrcccccccc}
    \setlength\tabcolsep{2.5pt}
    \tablecaption{uv-visibility Fitting Results}
    \tablehead{
\colhead{Name} & \colhead{} & \colhead{RA} & \colhead{$\delta$} & \colhead{Separation} & \colhead{$\sigma_{maj}$} & \colhead{$\sigma_{min}$} & \colhead{PA} & \colhead{\textit{i}} & \colhead{T$_{B}$} & \colhead{Int. Intensity}\\
\colhead{} & \colhead{} & \colhead{(J2000)} & \colhead{(J2000)} & \colhead{(au)} & \colhead{(mas)} & \colhead{(mas)} & \colhead{($^{\circ}$)} & \colhead{($^{\circ}$)} & \colhead{(K)} & \colhead{(mJy)}
}
    \startdata
L1448 IRS1 & -A & +03:25:09.455$_{-9e-05}^{+9e-05}$ & +30:46:21.83$_{-0.0001}^{+0.0001}$ &  & 193$_{-0.3}^{+0.3}$ & 93$_{-0.2}^{+0.2}$ & 28.0$_{-0.1}^{+0.1}$ & 61.1$_{-0.1}^{+0.1}$ & 397.7 & 48.6$_{-0.08}^{+0.08}$\\
\nodata & -B & +03:25:09.418$_{-0.0003}^{+0.0004}$ & +30:46:20.53$_{-0.0003}^{+0.0003}$ & 416 & 59$_{-0.9}^{+0.9}$ & 41$_{-0.9}^{+1}$ & 120.0$_{-2}^{+2}$ & 46.3$_{-1.5}^{+1.7}$ & 28.9 & 3.8$_{-0.03}^{+0.03}$\\
Per-emb-2 & -A & +03:32:17.936$_{-0.001}^{+0.001}$ & +30:49:47.63$_{-0.001}^{+0.001}$ &  & 59$_{-3}^{+3}$ & 50$_{-3}^{+3}$ & 66.0$_{-10}^{+10}$ & 32.1$_{-6.2}^{+5.8}$ & 35.1 & 2.5$_{-0.1}^{+0.1}$\\
\nodata & -B & +03:32:17.932$_{-0.002}^{+0.002}$ & +30:49:47.70$_{-0.002}^{+0.002}$ & 25 & 40$_{-5}^{+5}$ & 18$_{-7}^{+6}$ & 64.0$_{-10}^{+10}$ & 63.2$_{-11.1}^{+10.7}$ & 11.2 & 0.8$_{-0.06}^{+0.06}$\\
\nodata & -C & +03:32:17.935$_{-0.001}^{+0.001}$ & +30:49:47.27$_{-0.001}^{+0.001}$ & 109 & 117$_{-4}^{+4}$ & 67$_{-4}^{+4}$ & 126.0$_{-3}^{+3}$ & 54.9$_{-2.1}^{+2.2}$ & 5.7 & 4.7$_{-0.3}^{+0.3}$\\
\nodata & -D & +03:32:17.947$_{-0.001}^{+0.001}$ & +30:49:46.28$_{-0.001}^{+0.001}$ & 408 & 36$_{-3}^{+3}$ & 5$_{-2}^{+2}$ & 122.0$_{-5}^{+5}$ & 80.9$_{-3.2}^{+2.6}$ & 12.4 & 0.9$_{-0.04}^{+0.04}$\\
\nodata & -E & +03:32:17.920$_{-0.002}^{+0.002}$ & +30:49:48.17$_{-0.002}^{+0.002}$ & 174 & 64$_{-6}^{+6}$ & 42$_{-8}^{+7}$ & 108.0$_{-10}^{+20}$ & 47.3$_{-9.2}^{+9.9}$ & 3.4 & 1.1$_{-0.09}^{+0.1}$\\
Per-emb-5 &  & +03:31:20.942$_{-0.0003}^{+0.0003}$ & +30:45:30.19$_{-0.0003}^{+0.0003}$ &  & 340$_{-0.8}^{+0.8}$ & 221$_{-0.5}^{+0.5}$ & 30.0$_{-0.2}^{+0.2}$ & 49.6$_{-0.1}^{+0.2}$ & 60.1 & 184.4$_{-0.4}^{+0.4}$\\
NGC1333 IRAS2A & -A & +03:28:55.575$_{-9e-05}^{+8e-05}$ & +31:14:36.92$_{-9e-05}^{+8e-05}$ &  & 85$_{-0.2}^{+0.2}$ & 71$_{-0.3}^{+0.3}$ & 112.0$_{-0.7}^{+0.7}$ & 33.8$_{-0.4}^{+0.4}$ & 291.9 & 92.5$_{-0.2}^{+0.2}$\\
\nodata & -B & +03:28:55.568$_{-0.0004}^{+0.0004}$ & +31:14:36.31$_{-0.0005}^{+0.0005}$ & 186 & 56$_{-2}^{+2}$ & 38$_{-0.9}^{+0.8}$ & 13.0$_{-3}^{+3}$ & 46.2$_{-2.4}^{+2.2}$ & 59.3 & 9.7$_{-0.1}^{+0.1}$\\
Per-emb-17 & -A & +03:27:39.112$_{-6e-05}^{+6e-05}$ & +30:13:02.97$_{-6e-05}^{+6e-05}$ &  & 50$_{-0.2}^{+0.2}$ & 37$_{-0.2}^{+0.2}$ & 120.0$_{-0.6}^{+0.6}$ & 41.9$_{-0.4}^{+0.4}$ & 135.6 & 16.5$_{-0.03}^{+0.03}$\\
\nodata & -B & +03:27:39.123$_{-0.0002}^{+0.0001}$ & +30:13:02.74$_{-0.0002}^{+0.0002}$ & 83 & 95$_{-0.7}^{+0.7}$ & 51$_{-0.5}^{+0.5}$ & 161.0$_{-0.5}^{+0.5}$ & 57.3$_{-0.4}^{+0.4}$ & 41.7 & 10.7$_{-0.05}^{+0.05}$\\
Per-emb-18+ & -A & +03:29:11.267$_{-0.0003}^{+0.0004}$ & +31:18:30.99$_{-0.0003}^{+0.0003}$ &  & 44$_{-0.8}^{+0.8}$ & 18$_{-1}^{+1}$ & 73.0$_{-2}^{+2}$ & 65.1$_{-2.0}^{+2.1}$ & 22.1 & 4.1$_{-0.05}^{+0.05}$\\
\nodata & -B & +03:29:11.260$_{-0.0004}^{+0.0005}$ & +31:18:30.97$_{-0.0003}^{+0.0003}$ & 25 & 61$_{-1}^{+1}$ & 14$_{-3}^{+2}$ & 80.0$_{-1}^{+1}$ & 76.8$_{-2.3}^{+2.4}$ & 15.1 & 4.3$_{-0.06}^{+0.06}$\\
Per-emb-21+ &  & +03:29:10.674$_{-0.0001}^{+0.0001}$ & +31:18:20.09$_{-0.0001}^{+0.0001}$ &  & 50$_{-0.3}^{+0.3}$ & 48$_{-0.4}^{+0.4}$ & 74.0$_{-8}^{+8}$ & 16.1$_{-2.5}^{+2.1}$ & 87.4 & 16.3$_{-0.05}^{+0.05}$\\
Per-emb-22 & -A & +03:25:22.417$_{-5e-05}^{+5e-05}$ & +30:45:13.16$_{-5e-05}^{+5e-05}$ &  & 49$_{-0.2}^{+0.2}$ & 32$_{-0.2}^{+0.1}$ & 39.0$_{-0.5}^{+0.4}$ & 49.9$_{-0.3}^{+0.3}$ & 141.5 & 13.9$_{-0.02}^{+0.02}$\\
\nodata & -B & +03:25:22.359$_{-0.0001}^{+0.0001}$ & +30:45:13.07$_{-0.0001}^{+0.0001}$ & 227 & 41$_{-0.4}^{+0.5}$ & 23$_{-0.5}^{+0.5}$ & 19.0$_{-0.8}^{+0.8}$ & 56.2$_{-1.0}^{+1.0}$ & 102.1 & 5.4$_{-0.02}^{+0.02}$\\
L1448 IRS3B+ & -A & +03:25:36.324$_{-0.0001}^{+0.0001}$ & +30:45:14.81$_{-0.0002}^{+0.0002}$ &  & 52$_{-0.9}^{+0.9}$ & 23$_{-0.3}^{+0.3}$ & 25.0$_{-0.7}^{+0.7}$ & 63.7$_{-0.6}^{+0.6}$ & 86.1 & 5.3$_{-0.04}^{+0.04}$\\
\nodata & -B & +03:25:36.319$_{-0.0001}^{+0.0002}$ & +30:45:15.06$_{-0.0002}^{+0.0002}$ & 78 & 100$_{-0.9}^{+0.8}$ & 63$_{-0.4}^{+0.4}$ & 12.0$_{-0.7}^{+0.6}$ & 51.0$_{-0.4}^{+0.4}$ & 61.2 & 13.9$_{-0.06}^{+0.06}$\\
\nodata & -C & +03:25:36.387$_{-0.0001}^{+0.0001}$ & +30:45:14.64$_{-0.0001}^{+0.0001}$ & 247 & 200$_{-0.3}^{+0.3}$ & 175$_{-0.3}^{+0.3}$ & 25.0$_{-0.4}^{+0.4}$ & 28.7$_{-0.2}^{+0.2}$ & 56.1 & 78.3$_{-0.1}^{+0.1}$\\
L1448 IRS3A+ &  & +03:25:36.506$_{-0.0002}^{+0.0002}$ & +30:45:21.80$_{-0.0001}^{+0.0002}$ &  & 363$_{-0.2}^{+0.2}$ & 142$_{-0.2}^{+0.2}$ & 142.0$_{-0.07}^{+0.07}$ & 49.1$_{-0.1}^{+0.1}$ & 11.4 & 75.4$_{-0.1}^{+0.09}$\\
L1448 IRS3C & -A & +03:25:35.675$_{-8e-05}^{+8e-05}$ & +30:45:34.02$_{-9e-05}^{+9e-05}$ &  & 134$_{-0.2}^{+0.3}$ & 58$_{-0.2}^{+0.2}$ & 38.0$_{-0.1}^{+0.1}$ & 64.5$_{-0.1}^{+0.1}$ & 55.3 & 35.2$_{-0.06}^{+0.05}$\\
\nodata & -B & +03:25:35.679$_{-9e-05}^{+8e-05}$ & +30:45:34.26$_{-0.0001}^{+0.0001}$ & 72 & 66$_{-0.3}^{+0.3}$ & 30$_{-0.2}^{+0.2}$ & 38.0$_{-0.3}^{+0.3}$ & 62.5$_{-0.2}^{+0.2}$ & 69.3 & 11.9$_{-0.03}^{+0.03}$\\
Per-emb-35 & -A & +03:28:37.097$_{-0.0001}^{+0.0001}$ & +31:13:30.71$_{-0.0001}^{+0.0001}$ &  & 81$_{-0.4}^{+0.4}$ & 34$_{-0.2}^{+0.2}$ & 32.0$_{-0.2}^{+0.2}$ & 65.3$_{-0.2}^{+0.2}$ & 81.0 & 15.1$_{-0.04}^{+0.04}$\\
\nodata & -B & +03:28:37.225$_{-0.0008}^{+0.0008}$ & +31:13:31.67$_{-0.0007}^{+0.0007}$ & 570 & 113$_{-3}^{+3}$ & 18$_{-1}^{+1}$ & 48.0$_{-0.7}^{+0.7}$ & 80.8$_{-0.6}^{+0.6}$ & 14.1 & 2.9$_{-0.04}^{+0.05}$\\
NGC1333 IRAS2B & -A & +03:28:57.379$_{-0.0001}^{+0.0001}$ & +31:14:15.67$_{-9e-05}^{+0.0001}$ &  & 162$_{-0.3}^{+0.3}$ & 97$_{-0.3}^{+0.3}$ & 104.0$_{-0.2}^{+0.2}$ & 53.3$_{-0.2}^{+0.2}$ & 122.0 & 97.3$_{-0.1}^{+0.2}$\\
\nodata & -B & +03:28:57.373$_{-0.0003}^{+0.0004}$ & +31:14:15.98$_{-0.0003}^{+0.0004}$ & 95 & 50$_{-0.9}^{+0.8}$ & 43$_{-1}^{+1}$ & 92.0$_{-6}^{+6}$ & 29.3$_{-4.4}^{+3.8}$ & 48.4 & 8.9$_{-0.07}^{+0.08}$\\
SVS13A+ & -A & +03:29:03.772$_{-0.0002}^{+0.0002}$ & +31:16:03.71$_{-0.0002}^{+0.0001}$ &  & 72$_{-0.4}^{+0.4}$ & 68$_{-0.5}^{+0.5}$ & 80.0$_{-4}^{+4}$ & 20.3$_{-1.5}^{+1.4}$ & 209.4 & 59.9$_{-0.2}^{+0.2}$\\
\nodata & -B & +03:29:03.748$_{-0.0003}^{+0.0003}$ & +31:16:03.73$_{-0.0002}^{+0.0002}$ & 92 & 141$_{-0.7}^{+0.6}$ & 115$_{-0.6}^{+0.7}$ & 74.0$_{-0.9}^{+0.9}$ & 35.8$_{-0.5}^{+0.5}$ & 105.6 & 87.6$_{-0.4}^{+0.4}$\\
SVS13A2+ &  & +03:29:03.392$_{-0.0006}^{+0.0006}$ & +31:16:01.52$_{-0.0006}^{+0.0008}$ &  & 54$_{-2}^{+2}$ & 36$_{-1}^{+1}$ & 26.0$_{-4}^{+5}$ & 47.6$_{-3.2}^{+2.9}$ & 64.6 & 9.4$_{-0.2}^{+0.2}$\\
SVS13B+ &  & +03:29:03.085$_{-0.002}^{+0.002}$ & +31:15:51.64$_{-0.001}^{+0.001}$ &  & 302$_{-4}^{+4}$ & 180$_{-3}^{+4}$ & 83.0$_{-1}^{+1}$ & 53.3$_{-0.9}^{+0.9}$ & 36.8 & 48.1$_{-0.8}^{+0.8}$\\
\enddata
    \tablecomments{The separations are given in units of au, which is derived based on the average distance to the Perseus molecular cloud of 300~pc. Companion separations are defined as the distance from the first target listed.}
    \tablecomments{The \textit{uv}-plane nested sampling results. The models fit multiple Gaussians to the \textit{uv} data rather than the image-plane, which is sensitive to image reconstruction efforts and the specific beam shape, particularly towards marginally-resolved or un-resolved sources.}\label{table:13mmuvfits}
    \end{deluxetable}
    \end{longrotatetable}

\section{Results}
\begin{figure}
    \centering
    \includegraphics[width=\textwidth]{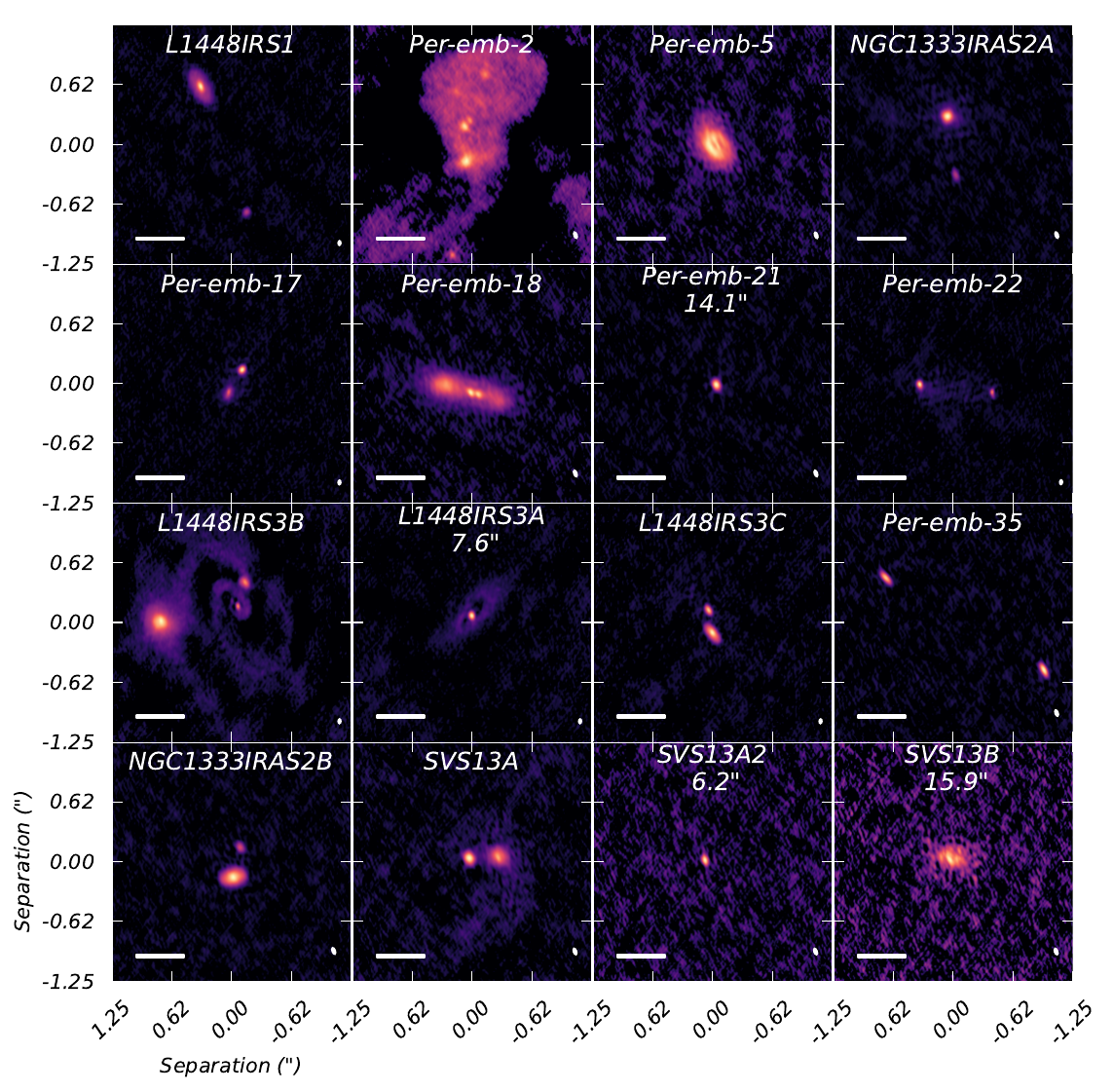}
    \caption{ALMA 1.3~mm continuum images of the Perseus multiples, constructed with a Briggs robust weighting parameter of 1. The sources are contained within 12 pointings detailed in Table~\ref{table:13mmpointing}. The distance given underneath the source name indicates the distances of that companion to the primary pointing source (the prior source in the list). If no distance is given, the center of the image is near to the center of the primary pointing. A 0\farcs5~(150~au)~scalebar is shown in the lower left, and the respective restoring beam is shown in the lower right. The colorscale is square-root scaled, with the lower bound set by a common RMS value of 20~$\mu$Jy~\beam}\label{fig:cont}
\end{figure}

With these observations, we detected all the circumstellar disks toward each multiple system within the survey at \ab8~au\space resolution for the 1.3~mm observations and \ab26~au\space resolution for the 3~mm observations, and most circumstellar disks are at least marginally-resolved. The ALMA images are shown in Figure~\ref{fig:cont} with the respective beam sizes in the lower-right corner. While we resolved out much of the $>$100~au~ ($>$0\farcs3) scale disk structures previously resolved \citep[][]{2018tobinlooneyli}, we did recover a large variety of disk sub-structures never previously resolved toward these sources. We briefly detail some key observations here and further discuss the morphologies of the individual sources in Appendix~\ref{sec:specificsources}.

\begin{figure}
    \centering
    \includegraphics[width=\textwidth]{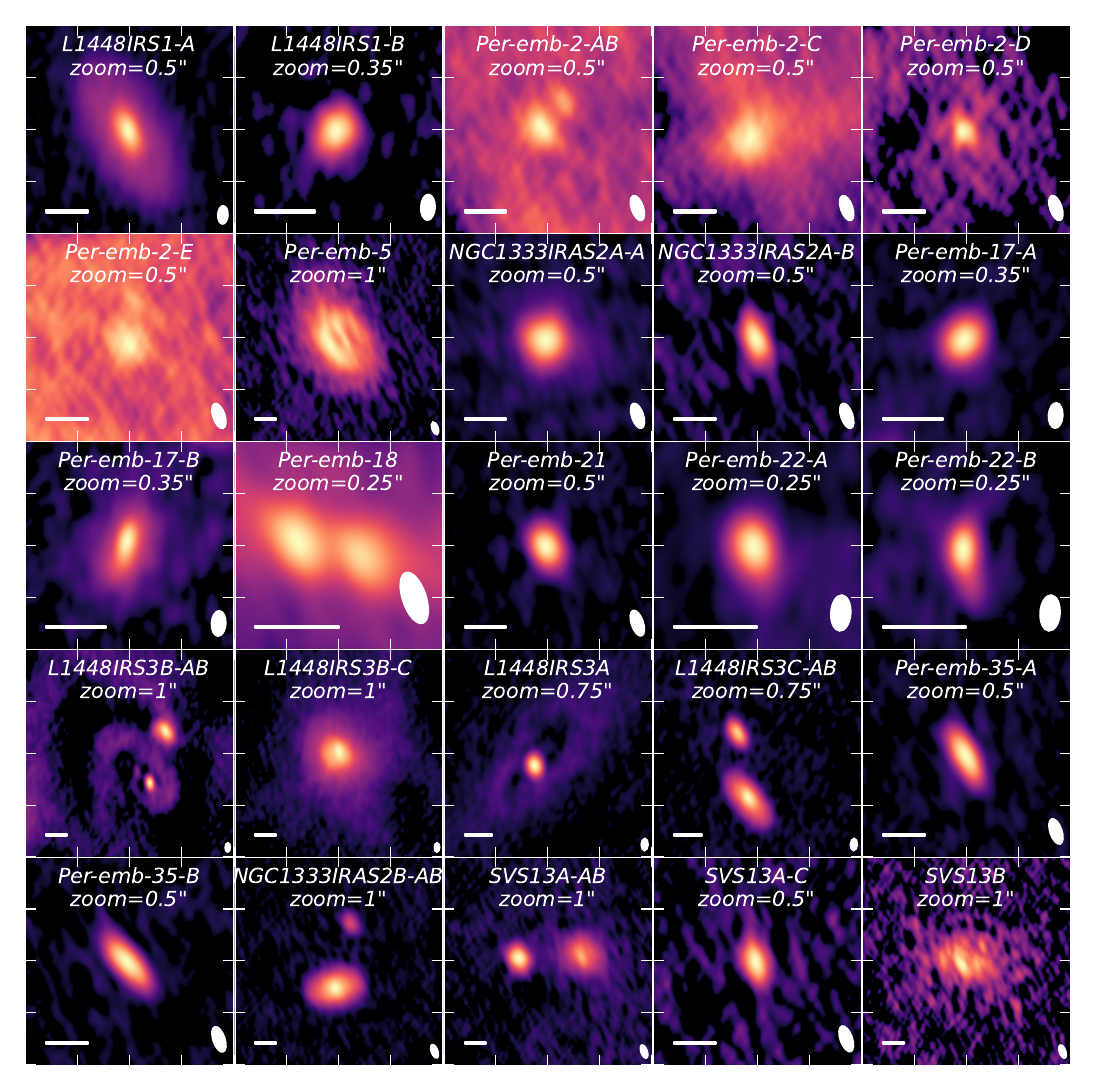}
    \caption{Similar image to Figure~\ref{fig:cont} but zoomed in to each fitted source in Table~\ref{table:13mmuvfits}. The box size for each plot is given and denotes the width and height of the plot. A 0\farcs1~(30~au)~scalebar is shown in the lower left and the respective restoring beam is shown in the lower right.}\label{fig:contzoom}
\end{figure}
\begin{figure}
    \centering
    \includegraphics[width=\textwidth]{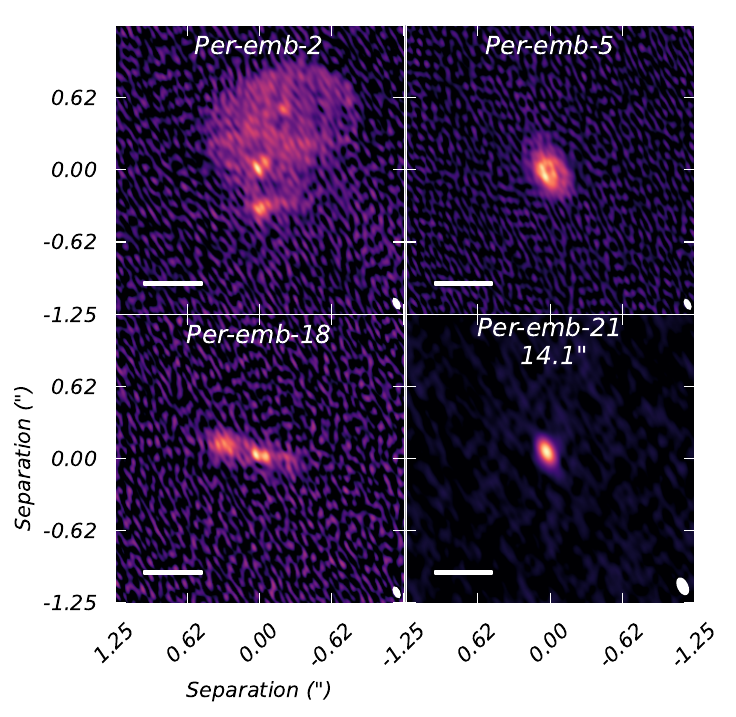}
    \caption{Continuum images at 3~mm of the Perseus multiples. Per-emb-2, Per-emb-5, and Per-emb-18 are constructed with superuniform weighting with an average restoring beam of 0\farcs09$\times$0\farcs04; while Per-emb-21 is constructed with Briggs robust 0.5, having a restoring beam of 0\farcs11$\times$0\farcs05, in order to recover the source from the noise. The sources are contained within 3 pointings detailed in Table~\ref{table:3mmpointings}.  The distance given underneath the source name indicates the distances of that companion to the primary pointing source (the prior source in the list). If no distance is given, the center of the image is near to the center of the primary pointing. A 0\farcs5~(150~au)~scalebar is shown in the lower left and the respective restoring beam is shown in the lower right.}\label{fig:3mmcont}
\end{figure}
\begin{figure}
    \centering
    \includegraphics[width=\textwidth]{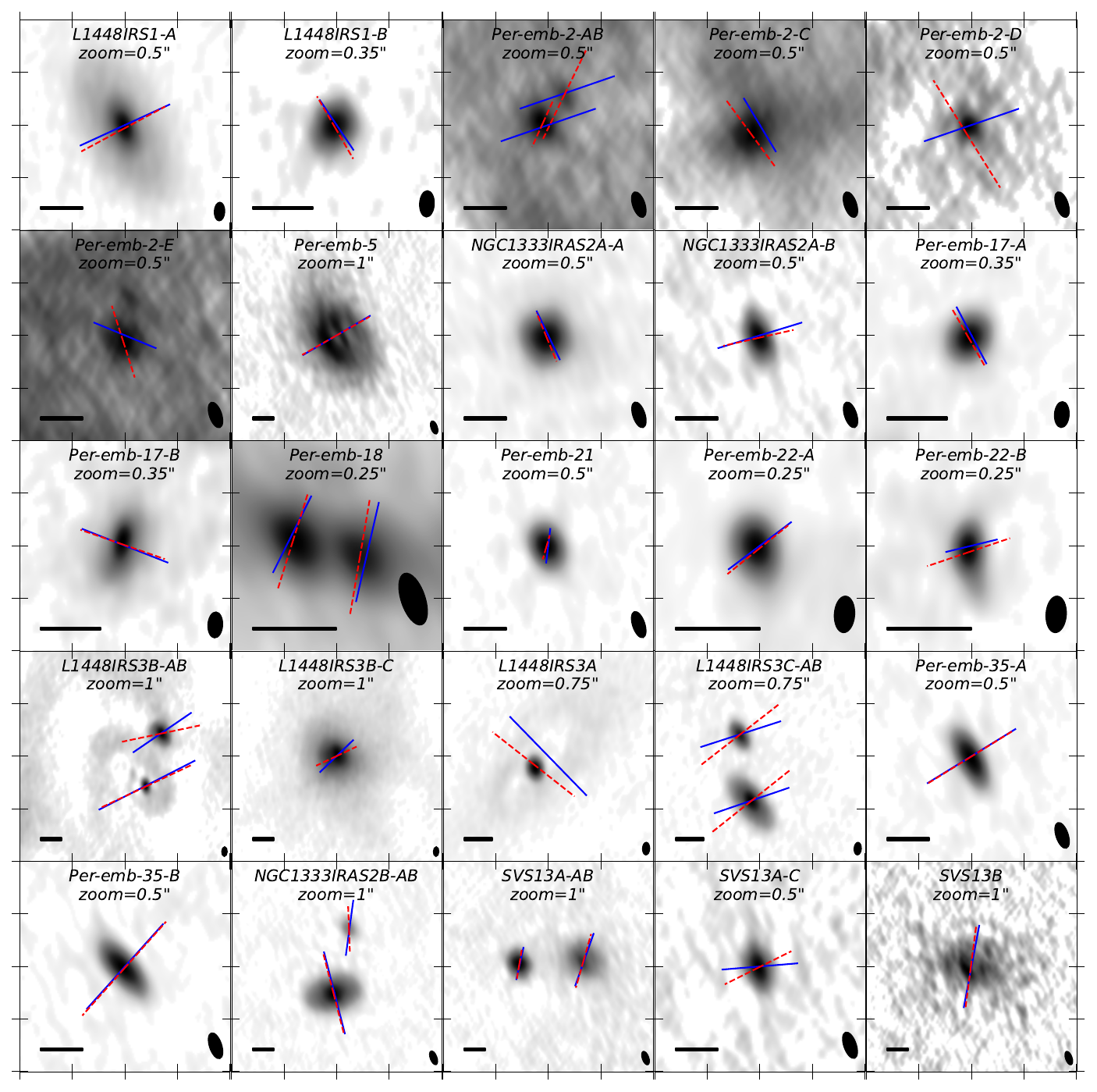}
    \caption{Summary of the source fitting technique overlaid on the same image as Figure~\ref{fig:contzoom}. The orientation vectors (angles orthogonal to the disk major axis) are indicated as colored lines for each of the fit sources\deleted{ and are length-scaled to the inclination of the disk, such that more inclined disks have a ``longer'' length vector}. The blue line indicates the image-plane derived orientation vectors while the dashed red line indicates the \textit{uv}-visibility derived orientation vectors. Several sources have orientation vector centers that do not align between the two fitting techniques. This is likely due to the image-plane fitting is influenced by the larger scale emission structure that is present such as Per-emb-2-C/-D, while the \textit{uv}-visibilities are selected to remove spatial scales larger than 0\farcs5 from the fit. \added{We adopt the \textit{uv}-visibility fit for the purpose of the analysis and note the \textit{image}-plane analysis are entirely consistent with the presented results.} }\label{fig:jveczoom}
\end{figure}

\subsection*{Per-emb-2}\label{sec:per2}
\begin{figure}
    \centering
    \includegraphics{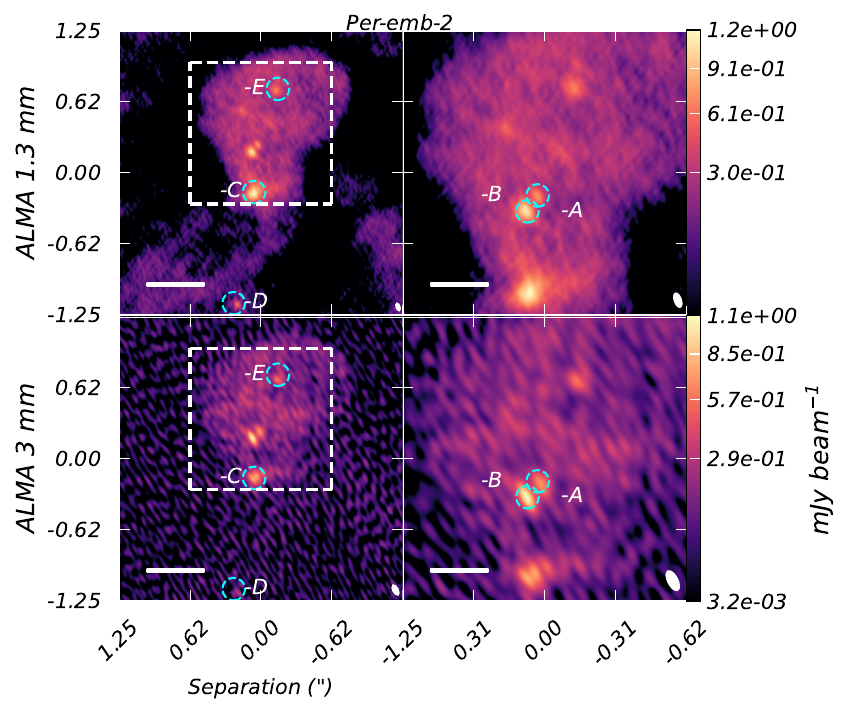}
    \caption{Left side is a 2\farcs5\space image of the Per-emb-2 system whose inner binary has been resolved. The right side is a 2$\times$\space zoom in on the source. The top row is the ALMA 1.3~mm observations and the bottom row is the ALMA 3~mm observations. In both observations, we resolve the compact inner binary but also report potentially 3 additional companions, 2 more within the disk and the southern-most companion falls just on the edge of the image. This would make Per-emb-2 a possible 5 companion protostellar system. A 0\farcs5~(150~au)~scalebar is shown in the left panels and a 0\farcs25~scalebar in the right panels. The beam is located in the lower right of both images. The colormap is square-root scaled.}
    \label{fig:per2}
\end{figure}

This source was previously reported to be a close multiple $<$50~au with the VLA at 9~mm \citep[][]{2016tobinlooney} and was further observed in \citet{2015tobinlooneywilner,2018tobinlooneyli} as a smooth continuum surface brightness distribution at 1.3~mm (Figure~\ref{fig:per2}). We resolved the compact binary (a\ab80~mas$\approx24$~au) in both the 1.3~mm and 3~mm observations. Our observations resolved much of the large scale emission, but further revealed a possible additional 3 compact sources with separations of \ab0\farcs431~(\ab129~au), \ab1\farcs432~(\ab430~au), and \ab0\farcs5~(\ab150~au) for the southern (SNR$\approx$25), northern-most (SNR$\approx$22), and southern-most (SNR$\approx$15) sources respectively, relative to the Per-emb-2-A source of the compact binary. All of these compact sources (except the southern-most companion) are found within regions of relatively enhanced surface brightness within the \citet{2018tobinlooneyli}\space observations, presumed to be a massive extended disk. The sources are present at least the 5~$\sigma$\space level in the 1.3~mm and the 3~mm observations. It should be noted that only the compact sources -A,-B  and a marginal detection of the brighter southern source -C appear in 9~mm VLA observations \citep[][]{2016tobinlooney}\space whereas more diffuse sources, the northern-most -D and and southern-most -E sources do not.

\subsection*{Per-emb-5}\label{sec:per5}
\begin{figure}
    \centering
    \includegraphics{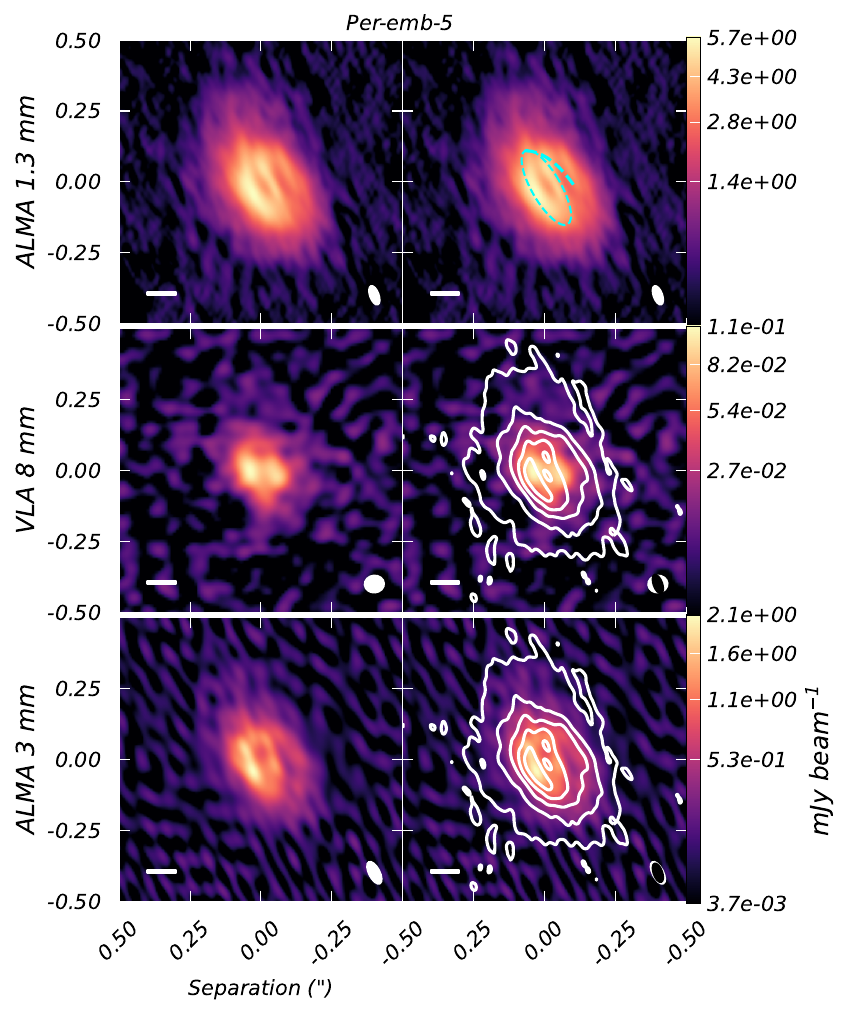}
    \caption{Multiwavelength observations of the Per-emb-5 system. The upper panels are naturally weighted clean images from the 1.3~mm ALMA data. The middle panels are from \cite{2016tobinlooney} robust 0.25 VLA 9~mm. The bottom panels are super-uniform clean images from the 3~mm ALMA data set reported here. The disk is asymmetric and appears as a ring and a single-arm spiral structure. We overlay an ellipse and line to assist in defining the substructure of Per-emb-5. A 0\farcs1~(30~au)~scalebar is shown in the lower left and the representative beam is in the lower right. The white contours start at the 5~$\sigma$~level and iterate by 20~$\sigma$, where the 1~$\sigma=50$~$\mu$Jy~\beam. The colormap is square-root scaled.}
    \label{fig:per5}
\end{figure}

Per-emb-5 was also previously reported to be a close multiple $<$50~au with the VLA at 9~mm \citep[][]{2016tobinlooney}. We instead found continuum emission that appears consistent with that of a disk surrounding what appeared as two peaks in the VLA data (Figure~\ref{fig:per5}).  The disk appeared to have a central cavity (centered between the two VLA peaks), and maybe a single spiral arm to the west. There is an asymmetry across the minor axis of the disk and a flux enhancement in the south-east portion of the disk.

\subsection*{L1448~IRS3B}\label{sec:irs3b}
\begin{figure*}[htb!]
    \centering
    \includegraphics[height=7in]{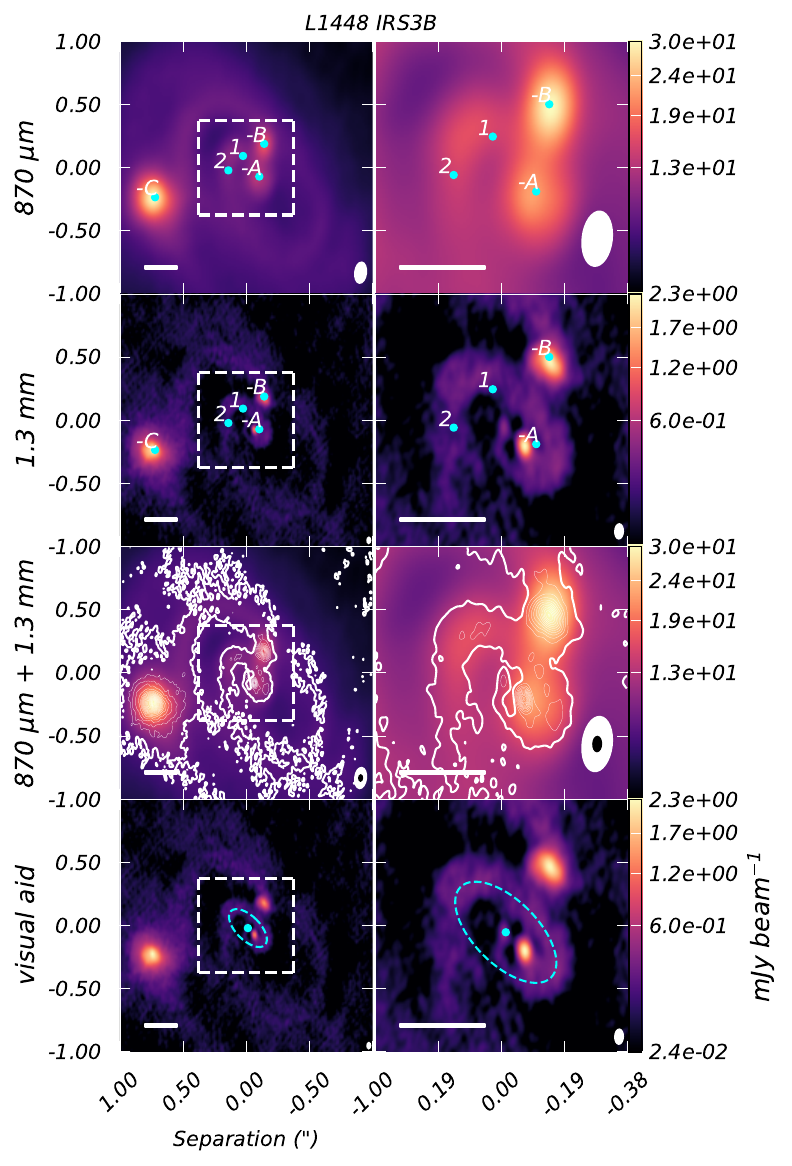}
    \caption{The left side is the 2\arcsec\space view of the system L1448~IRS3B, in which you can clearly see the triple  system and the spiral arm substructures. Right side is a \ab4$\times$~zoom in on the inner disk of the binary L1448~IRS3B-AB. The top row is reproduced from \cite{2021reynoldstran}\space using ALMA 870~\micron\space continuum emission. The upper-middle row is the current 1.3~mm, 8~au resolution observations.  The lower-middle row has the 870~\micron\space data as the background and the 1.3~mm data is overlayed as white contours. The lower row is the same 1.3~mm data but with diagrams to aid in visualizing the likely configuration of the disk. We now resolve the inner disk to be a ring and the so called ``clump'' \citep[see Figure 2; ][]{2021reynoldstran}, is the north-east side of the ring. We report the L1448~IRS3B-A source is now resolved as just inside of the ring and L1448~IRS3B-B is now resolved as just outside of the ring. The separation between the two sources is well resolved and the L1448~IRS3B-B source is resolved. Towards the geometric center of the ring there is compact emission. We also reproduce the kinematic centers as the numbers ``1'' and ``2'' in the plot, as reported in \cite{2021reynoldstran}. The 0\farcs25~(75~au)~scalebar is shown in the lower left and the representative beam is in the lower right. The white contours start at the 3~$\sigma$~level and iterate by 10~$\sigma$, where the 1~$\sigma=20$~$\mu$Jy~beam$^{-1}$. The colormap is square-root scaled. The contour representative beam is overlayed in black at the lower right.}
    \label{fig:l1448irs3b}
\end{figure*}

L1448 IRS3B is certainly a exceptional source (Figure~\ref{fig:l1448irs3b}). The system is home to at least 4 compact continuum sources within 8\arcsec, 3 of which are within 1\arcsec\space of each other. The brightest feature, the tertiary companion commonly known as L1448~IRS3B-C is likely optically thick. The tertiary companion is embedded within one of the large spiral arms that stems from the inner disk to the outer disk. Zooming into the center of the system, two bright continuum sources are obvious, one just inside of an inner spiral arm/disk structure and one just outside. We apparently resolved the ``clump'' as reported in \citet{2021reynoldstran} as the north-east portion of the inner disk and now report an additional faint compact source near the geometric center of the inner disk. The bright point L1448~IRS3B-A is now resolved as the south-west portion of the ring and a bright source just inside of the ``ring'' of the inner disk. L1448~IRS3B-B is just outside of the ``ring''.


We reproduced Figure 16 from \citet{2021reynoldstran} in Figure~\ref{fig:l1448irs3b}, denoting the locations of the kinematic centers for the L1448~IRS3B system using various disk tracing molecular line observations and techniques. We visually depict the new geometric center of the inner ring, which coincides with the center of the ``deficit'' previously reported and overlaps, within observation uncertainties, with the center-point of the kinematic centers. It is possible this newly resolved source is another deeply embedded protostellar source and the disk could harbor as many as 4 protostellar sources. For the purpose of the analysis conducted later in the paper, we do not consider this small point source at the center as another independent source; instead, we only consider the two confirmed compact continuum sources as protostar sources. The new source, while confidently detected, is too faint to have its geometric parameters well-constrained from Gaussian fitting. We refer to this new source as L1448 IRS3B-D, centered  at 03h25m36.326s~30:45:14.93.
The designation of sources in L1448 IRS3B may need reassignment in the future once the nature of the source and the inner disk are better characterized with additional observations.

\subsection*{L1448 IRS3A}
\begin{figure}
    \centering
    \includegraphics{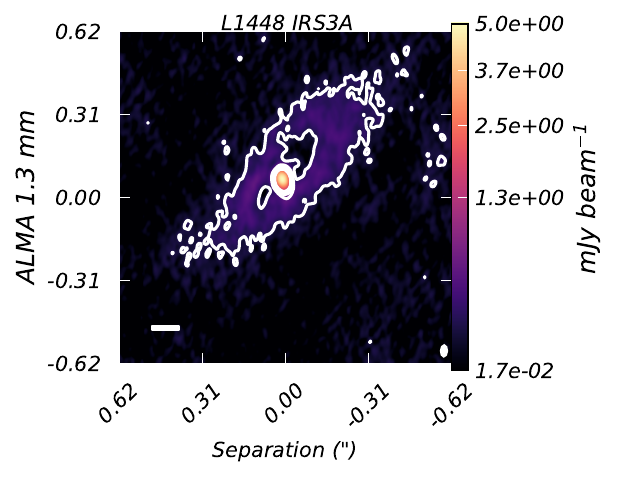}
    \caption{The source L1448 IRS3A is within the ALMA primary beam centered on L1448 IRS3B. We resolve the circumstellar disk to be a ring, surrounding the bright compact source. The data is reconstructed using a Briggs robust 1. A 0\farcs1~(30~au)~scalebar is shown in the lower left and the representative beam is in the lower right. The white contours start at the 5~$\sigma$~level and iterate by 20~$\sigma$, where the 1~$\sigma=17$~$\mu$Jy~\beam. The colormap is linear scaled.}
    \label{fig:l1448irs3a}
\end{figure}

Within the same field-of-view as the L1448 IRS3B observations, we resolved the disk around L1448 IRS3A (Figure~\ref{fig:l1448irs3a}). Previous ALMA 870~\micron\space observations hinted of substructure in the surrounding circumstellar disk of L1448 IRS3A, but were unable to determine the nature of the features \citep{2021reynoldstran}. Additionally, \citet{2021reynoldstran}\space determined the structure was likely not spiral arms due to the disk being relatively stable against collapse. With the higher resolutions afforded by these observations, we resolve the circumstellar material to be a ring surrounding a compact source which remains unresolved in these observations. The compact continuum source appears slightly off the geometric center of the ring, but this could be explain by projection effects of an inclined disk on the observations.

\section{Statistical Analysis of Orientations}\label{sec:statanalysis}

\subsection{Companion Finding}
In order to facilitate the analysis of the relative disk orientations in each multiple system, we must first determine which systems are associated with each other. While companions $<$100~au are mostly trivial to assign, systems with $>$3 sources and separations that range out to \ab10,000~au require a more automated approach. We made use of a method similar to the one implemented by
\citet{2022tobinoffnerkratter} which will automatically create the companion associations \added{by calculating the line-of-sight (L.O.S.) distances} given an input catalog of positions.\added{ The systems constructed by this algorithm are verified to be consistent with prior studies \citep{2016tobinlooney,2018tobinlooneyli} and the algorithm is discussed in detail in Appendix~\ref{sec:companionfinding}.}

\subsection{Geometric Orientations}
To investigate the proto-multiple disk orientation alignments, we measured the projected angular difference of the inclination and position angle of the disks. With the full 2-D Gaussian fit parameters provided from the CASA \textit{imfit}\space and \textit{uv}-visibility fitting, we constructed the orientation vectors for each of the sources. We assumed that the disks are axisymmetric and geometrically thin and solved for the inclination of the disk from the major and minor axis lengths, deconvolved from the synthesized beam, via arccos($\frac{\sigma_{minor}}{\sigma{major}}$). The position angle of the disk is directly output from the 2-D Gaussian fits as the angle of the major axis, deconvolved from the synthesized beam, oriented to the standard east-of-north.

From a single continuum observation alone, we are insensitive to the orbital motions of the disk, thus we were unable to differentiate between aligned and anti-aligned disks. So for our analysis, we restricted \textit{i}\space to range from 0-90\deg\space(such that face-on is 0\deg) and the PA to range from 0-180\deg. The resulting dot product of the angular momentum vectors is normalized to fall between 0-1, and we constructed the summary plot in Figure~\ref{fig:jveczoom}\space by overlaying resulting orientation vectors on the continuum image from Figure~\ref{fig:cont}. In most  sources the image-plane and the \textit{uv}-visibility derived orientation vectors are nearly aligned, however for the sources where the compact disks are either unresolved or marginally resolved, the methods yield slightly different vectors. In these cases, for the purpose of interpretation, we favored the \textit{uv}-visibility derived results as these results will take full advantage of the spatial-dynamic range in the data. We present the image-plane fits in Appendix~\ref{sec:imagefit}\space and the results of the statistical tests in Appendix~\ref{sec:comparisonorientationvectors}.

\subsection{Models of Companion Orientations}\label{sec:models}
Our goal is to statistically analyze the companion orientations within the sample and provide a robust way to characterize the ensemble sample of orientations. To do so, we need to generate a sample of model protostellar configurations to determine the most probable formation pathways for the observed sample. The protostellar configurations we considered for the analysis are disks that are preferentially aligned with each other, representing the expectation of companions formed via disk fragmentation, and randomly aligned disks as would be expected to result from turbulent fragmentation.

To generate these configurations, the parameters we considered to generate a single system are: stellar multiplicity\deleted{, the separation of the companions, and}, the position angle, and inclination of the compact circumstellar disks (ignoring circum-multiple disks and under the assumption the stellar angular momentum axis and disk angular momentum axis are aligned)\deleted{, and the size of the disk (enforcing disks around sources are smaller than the separation to the nearest companion)}. From these parameters alone, we constructed our model distributions. To model an empirically-driven distribution of proto-multiple systems, we sampled multiplicity for separations between 20-10,000~au, following the distribution of Perseus protostellar multiples of \citet{2022tobinoffnerkratter}.

From these constructed systems, we generated a range of possible protostellar configurations with various fractions of aligned and randomly-aligned orientations. That is to say, a system with a multiplicity of 4 sources and a fractional orientation of 75\%\space preferentially aligned sources and 25\%\space randomly aligned sources, will have on average, 3 preferentially aligned sources and 1 randomly aligned source. We inserted no bias or preferential weighting corresponding to the separations of the sources with regards to alignment.

\added{Preferentially aligned systems are constructed by drawing the position angle and inclination from a normal distribution described by a full-width-half-maximum value of 30\deg~\citep{2019leeoffner}. The choice of 30\deg\space is chosen to be consistent with similar studies carried out but the results presented are not significantly changed if moderate deviations ($\pm$5\deg) from 30\deg\space is chosen. The resulting distributions of predominantly preferentially aligned systems are similar to population synthesis simulations conducted by \citet{2018bate}. Randomly aligned systems are constructed by drawing the position angle and inclination from a uniform distribution of inverse cosine from 0 to 1.}

For a rigorous statistical analysis, the observations need to be compared against a continuous distribution of models. Thus we generated 10,000 model systems (each system will have at least two sources) for each of the different fractions of systems with fractional orientations, ranging from fully preferentially aligned to completely random alignments. When extracting the orientation vectors of the constructed systems, we applied the same 2D projection bias the empirical continuum observations are subject to (i.e. we normalized the p.a. to $<$180\deg\space and the inclination to $<$90\deg). This provides an observation-like set of models to compare directly with the observations. A visual representation of the median empirical cumulative distribution function (ECDF) for each of the constructed fractional alignment distributions is shown in Figure~\ref{fig:stats10010000500}, with the median ECDF of the observations with uncertainties is overlaid in black.

\subsection{Statistical Tests}\label{sec:statistics}
For the observational and model data, we evaluated the dot product of orientation vectors for unique disk pairs in every system (counting each disk pair only once; Table~\ref{table:13mmuvfits}). \added{We utilized an algorithm similar to the companion finding approach, where each compact binary within a system is compared first. We then randomly selected a source from each compact binary to use for further comparison. This gives us N$_{s} - 1$\space number of pairs per system where N$_{s}$\space is the number of sources within a given system (more detailed explanation in Appendix~\ref{sec:comparisonorientationvectors}). }\deleted{This gives us $\genfrac(){0pt}{2}{N_{s}}{2}$ number of pairs per system where N$_{s}$\space is the number of sources within a given system. }This resulting dot product is a derived from the results of the \deleted{\textit{imfit}\space and the }\textit{uv}-visibility fitting\added{ (\textit{imfit}\space analysis remains entirely consistent with the uv-visibility results and is detailed in Appendix~\ref{sec:imagefit})}. To account for the uncertainties in the observations and to resample the particular sources chosen for the pairs, we recalculated each of the dot products 10,000 times sampling the fit errors assuming Gaussian uncertainties. This gave a suite of 10,000 realizations of the empirical distribution which are all consistent with the observations within the uncertainties.

We need to construct a distribution of alignments that could form the basis for the observational underlying distribution. Since this underlying distribution is not known, we constructed a grid of distributions that would cover the range of possible alignment distributions. Each particular constructed distribution is an aggregate sample of preferentially aligned and randomly aligned distributions, sampled via some fractional ratio (e.g. C$_{0.75}$UC$_{0.25}=$3-to-1, aligned to randomly-aligned, etc). 

We then calculated the probability of each of the 10,000 resampled observed distributions being drawn with each of the fractional ratios (i.e. the null hypothesis, see Appendix~\ref{sec:stattest}) by utilizing the 2-sample Kolmogorov-Smirnov, Anderson-Darling, and Epps-Singleton probability tests \citep[][respectively]{andersondarling, kstest, eps1,eps2}. We set a null-hypothesis rejection threshold of 0.3\%~(3~$\sigma$), such that if the probability test can reject the null hypothesis at the 3~$\sigma$\space threshold, we discarded that particular empirical distribution. We then counted up all of the distributions that pass this threshold and summarize the full statistics in Table~\ref{table:stats10010000500uv}. A full description of the tests and methodologies is given in Appendix~\ref{sec:stattest}.

\added{To analyze the potential signatures of formation mechanism pathways, we divide the full set of observations into three subsamples and perform analysis on these subsamples. The subsamples chosen correspond to the maximum separation of companions, derived from the fit parameters: $<$100~au, $<$10,000~au, and $>$500~au. Surveys of protostellar multiplicity \citep[][]{2016tobinlooney, 2020tobinsheehanmegeath, 2021encaladalooney, 2022tobinoffnerkratter} found the average separation of companions to be \ab75~au, thus we chose 100~au as the compact subsample. Since turbulent fragmentation can form on 1000s~au scales and then migrate down to 100s~au scales, we select scales $>$500~au to select the subsample with a different underlying distribution. However, our selection of subsample noticeably foregoes the sources that fall within the range of separations $100<a<500$~au.  This selects out \ab13~source pairs. While performing such a selection cut reduces the overall number of sources included in the subsamples, the authors do not detect any major difference that would change the findings. This particular cut of $100<a<500$~au was chosen to ensure sources selected by the two subsamples $<100$~au and $>$500~au would probe different underlying distributions, minimizing the overlap in these particular distributions, making the resulting statistical test more sensitive to differences in the underlying distributions. }\deleted{While the full sample is also included in the primary text as the distribution with separations $<$10,000~au, we provide the remaining subsample for separations $<$500~au for completeness.}

\deleted{With the subsample of separations $<$500~au, the distribution tends towards the full sample distribution of separations $<$10,000~au and $>$500~au. This could be understood as the sources found between separations $100<a<500$~au are drawn from the same underlying distribution as the wider separation sources. Thus it is likely that for compact sources with separations $<$100~au, they are drawn from a distribution of preferentially aligned disk orientations within 30\deg\space while all sources $>$100~au are preferentially drawn from a randomly aligned distribution of orientations.}

When the sample is limited to $<$100~au\space separations (Left panel: Figure~\ref{fig:stats10010000500}), the statistical tests imply the fractional ratio of orientations is comprised primarily of preferentially aligned sources with distributions at least 40\%\space preferentially aligned.

When the sample is limited to only extended companions with separations $>$500~au\space separations (Right panel: Figure~\ref{fig:stats10010000500}), 15 source pairs are analyzed and we rule out distributions of fully preferentially aligned companions down to 80\%\space preferentially aligned. The results of the statistical results for each of the distributions appears equally likely for fractional alignments between 50\%-70\%\space preferentially aligned, thus we are not able to conclusively determine the exact alignment ratio.

Our analysis found the full observed sample (Middle panel: Figure~\ref{fig:stats10010000500}), separations out to 10,000~au, is most consistent with a hybrid population of multiples with some contribution from multiple systems with preferentially aligned orientation vectors (up to 80\% aligned sources) but is consistent with distributions down to 40\%\space fractional alignments. We strongly ruled out fully aligned and fully randomly aligned distributions of disk orientation pairs. While it would be expected for distributions of highly separated sources to consist of a higher fractional ratio of randomly aligned systems, we found this distribution is consistent with roughly an equal fractional ratio.

The results are not significantly different whether we used the result from the image-plane or \textit{uv}-visibility analysis, but the results from the \textit{uv}-plane analysis tend to be favored as the uncertainties in the fits are more constrained.

\section{Discussion}\label{sec:discussion}
\subsection{Formation Pathways}

There are two primary mechanisms for multiple star formation which operate at various scales, disk fragmentation and turbulent fragmentation during core collapse. Disk fragmentation operates on \ab100s~au scales in massive disks ($\frac{M_{d}}{M_{*}} \ga 0.1$) that are gravitationally unstable (i.e. Toomre~Q $<$~1), if the disk cools sufficiently fast \citep[][]{2001gammie}. The outcomes of gravitational instability can be observed directly \citep[such as; ][]{2016tobinkratter, 2021reynoldstran}\space but also has been extensively modeled \citep[][]{2006kratter, 2009boley, 2010krattermurray, 2013vorobyov, 2018vorobyov}. However clear cases of ongoing disk fragmentation are somewhat elusive aside from L1448~IRS3B, perhaps due to the short timescales before the disk self-stabilizes by redistributing the angular-momentum and/or fragments. Turbulent fragmentation typically operates on 1000s of au scales in turbulent cloud cores. Moreover, simulations show that systems may migrate significantly from their nascent locations \citep[][]{1999ostriker, 2019lee} due to gravitational attraction and initial velocities with respect to the cloud core. Stars in these environments can become bounded and unbounded with empirical evidence that suggests multiples frequently form via turbulent fragmentation \citep[][]{2016murillodishoeck}. Therefore, it is likely that gravitational instability produces companions with compact separations ($<100$~au), both pathways can produce companions with moderate separations ($<$500~au), and a single mechanism may primarily populate the more extended configurations at scales of $>$500-10,000s~of~au.


While we cannot directly infer the evolution of any one particular system, we can use statistical approaches to modeling and determine the most probable formation pathway for an ensemble of multiple systems. Particularly, \citet{2018bate}\space \added{conducted} hydrodynamic simulations of proto-multiples undergoing gravitational collapse in a viscous medium, and found for a set of unbiased sources, the relative alignment angles are not correlated with hierarchy number, separation distributions, or age \citep[see Figures 19 and 24 in ][]{2018bate}. Additionally \citet{2016offnerdunham} showed that systems formed through turbulent fragmentation are randomly aligned and found that partial misalignment persists even after inward orbital migration. This was further supported by \citet{2016leedunham} in observing companions with separations $a>$1,000~au who found a nearly complete random distribution of alignment angles. Analyzing our results of statistical tests (Table~\ref{table:stats10010000500uv}), we can statistically infer the fractions of preferentially aligned and randomly aligned orientation pairs that are in the given sample.

We found for our particular Perseus sample, for the distribution of companions with separations $<$10,000~au, distributions of a majority preferentially aligned orientation pairs are ruled out with at most 80\%\space preferentially aligned. The $a<$10,000~au sample also disfavors distributions of less than 40\%\space preferentially aligned sources. \deleted{The \textit{uv}-visibility derived tests have better constraints on the fitted uncertainties. }With the relatively low number of statistics (at most 21 pairs of sources), we were not able to determine the exact underlying fractional alignment ratio that best describes the sample of sources.\added{ This characterization of the subsamples and full sample is consistent with synthetic radiation hydrodynamical simulations of protostellar clusters \citep{2018bate}.}

\begin{figure}
    \centering
    \includegraphics[width=\textwidth]{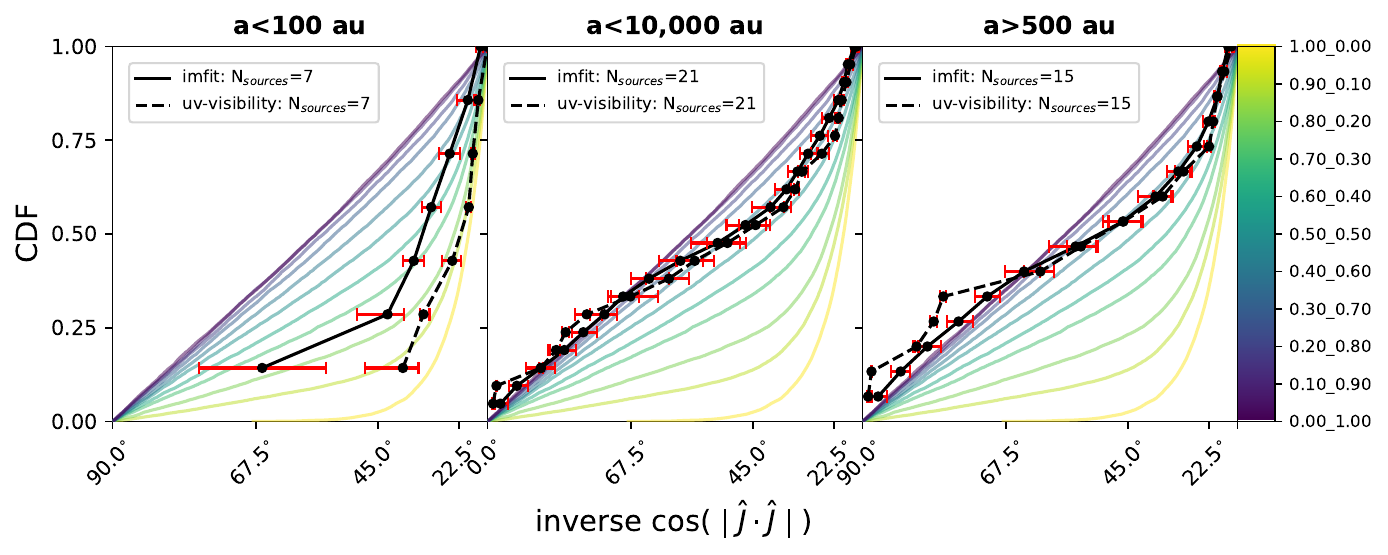}
    \caption{Left panel: corresponds to the subsample of compact companion separations ($<100$~au); Middle panel: the full companion sample with separations (a$<$10,000~au); Right panel: subsample of extended companions (a$>$500~au). Median CDFs of the dot-product of the orientation vectors the two fitting techniques and of the various distributions of fractional alignment ratio. Each of the colored lines corresponds to a particular sample of fractional ratio distributions between a fully preferentially aligned distribution to fully randomly aligned distribution, with the given ratio in the colorbar, as detailed in Section~\ref{sec:statistics}. We resample the data with Gaussian errors considering the observation uncertainties and construct empirical CDFs. The solid black line is a median CDF for the CASA \textit{imfit} results. The dashed black line is a median CDF for the \textit{uv}-plane fit results. The red, horizontal error bars associated with each ECDF, represent the 1~$\sigma$\space uncertainty for the lower and upper bound, respectively. The compact ($<$100~au) subsample is consistent with preferentially aligned distributions, with $>$80\%\space of all sources having a dot-product within 30\deg. The full sample and extended companion subsample are consistent with a ratio of aligned orientations, with $~$60\%\space of all sources having a dot-product within 30\deg.  The full statistical analysis is given in Table~\ref{table:stats10010000500uv}\space and further discussed in Appendix~\ref{sec:stattest}.}
    \label{fig:stats10010000500}
\end{figure}

It is likely the more compact companions of our sample ($a<$100~au, left panel of Figure~\ref{fig:stats10010000500}\space and Table~\ref{table:stats10010000500uv}), which do not rule out the null hypothesis in fractional ratio tests from 100\%\space preferentially aligned distributions down to 40\%\space preferentially aligned distributions, are formed primarily via gravitational instability, forming preferentially aligned pairs. This is similar to surveys of compact ($a\approx50$~au) star-planet binaries \citep{2022dupuykraus}\space which were observed to have mutual orbital inclinations $<$30\deg. The large spread in the fractional ratio tests is likely due to the low number of sources (n$=7$) that have separations $a<$100~au).\added{ \citet{2018bate} shows an EDCF of compact companions that appears to be most consistent with 80\%\space preferentially aligned sources.}

The extended samples ($a>$500~au, right panel of Figure~\ref{fig:stats10010000500}\space and Table~\ref{table:stats10010000500uv}) appear to reject distributions with more than 80\%\space preferentially aligned sources and distributions with less than  40\%\space preferentially aligned sources.

The full sample ($a<$10,000~au, middle panel of Figure~\ref{fig:stats10010000500}\space and Table~\ref{table:stats10010000500uv}) are likely more constrained due to the higher number of sources (n$=$21) and appear consistent with distributions of 80\%\space to 40\%\space preferentially aligned sources.

\subsection{Formation Mechanism of Individual Systems}
The statistical tests alone are not able to tell a complete picture of an individual system, but it can provide a statistical way to characterize a sample of systems. To best determine the formation mechanism of any particular system, a combination of multi-wavelength observations of continuum and molecular lines are needed. We would expect sources that have formed via gravitational instability to have separations on the order of the continuum disk size ($a\approx$100s~au) and the kinematics of the system to be organized. Whereas, formation of system via turbulent fragmentation happens on much larger scales ($a\approx$1,000-10,000s~au). \added{The sources formed at larger scales} could further migrate to \ab100s~of~au, preserving no preferential alignment of the kinematics. We discuss the likelihood of the observed systems deriving from these two formation mechanisms based on our 1.3~mm and 3~mm ALMA continuum observations combined with prior observations. 


\subsubsection{Disk Fragmentation Candidates}
We detected circum-multiple material in nine of the 11 observed sources (Per-emb-5 is now classified as a single source) in these observations (Figure~\ref{fig:contzoom}); however our array configuration is less sensitive to extended emission.  In all detected cases, the circum-multiple material was observed around Class 0 sources, while sources that show significant misalignment do not show much circum-multiple material within the detection limits of our observations. In selecting a subsample consisting of only Class 0 sources from our full sample of sources, we found the median relative orientation angle between the sources in each system is 24\deg\space for the Class 0 sources.

Seven of the observed multiple systems are consistent with having their $<$500~au companions formed via disk fragmentation. The sources Per-emb-2, Per-emb-17, Per-emb-18, Per-emb-22, L1448 IRS3B, L1448 IRS3C, and SVS 13A appear most consistent with the disk fragmentation formation pathway.

In particular for Per-emb-2, \citet{2020pinedaseguracox} found the mass-infall rate exceeds that of the accretion rate derived from the bolometric luminosity. This scenario provides favorable conditions to trigger gravitational instabilities that lead to the fragmentation of the disk \citep{2010krattermatzner}.

These compact circumstellar disks appear to be strongly aligned and deeply embedded Class 0 systems, with bright circum-multiple disks (in some cases).

\subsubsection{Turbulent Fragmentation Candidates}
There are several other systems with separations $a<$1,000~au which are not consistent with disk fragmentation and likely formed via a combination of other methods. The sources L1448 IRS1 and NGC1333 IRAS2A are relatively compact proto-multiple systems, with strongly misaligned disks.

The companion ($a\approx418$~au) of L1448 IRS1 is nearly orthogonal to the brighter primary source. This source could have formed via turbulent fragmentation and then evolved via dynamical interactions to more compact scales  (a\ab1\farcs4$\approx$420~au).

The companion of NGC1333 IRAS2A is nearly orthogonal to the brighter primary source with a much higher inclination within the limits of our observations. Our observations for the disk position angle are consistent with an angle orthogonal to prior observations of the outflow position angles \citep{2015tobindunhamlooney}. Similar to L1448 IRS1, this source could have formed via turbulent fragmentation and then evolved via dynamical interactions to more compact scales (a\ab617~mas$\approx$185~au).

Per-emb-35 appears to be two similar sources, with aligned relative disk orientations, at a wide separation of 1\farcs9~(\ab570~au). While the orientations are aligned with an average relative orientation angle of \ab19\deg, the source is not likely to have formed via gravitational instability. We report no detection of circumbinary material or any dust continuum between the two sources. This coupled with the wide separation means Per-emb-35 is likely to have formed via large-scale (1000s~of au) mechanisms, likely turbulent fragmentation or dynamical capture, and migrated to more compact scales (100s~of au).

The sources Per-emb-21, L1448 IRS3A, and SVS13 B are wide ($a>1000$~au) companion sources and additionally appear to not be preferentially aligned towards the outer sources in the respective systems, and are thus likely formed via turbulent fragmentation.

\subsubsection{Ambiguous Formation}
It is not readily apparent via which formation pathway NGC 1333 IRAS2B ($a\approx95$~au) formed. The orientation vectors are relatively aligned within 40\deg\space but this could be happenstance as there is no circum-multiple material found in these or prior observations. The system could have formed at long separations and migrated to compact scales.

\begin{deluxetable}{lrrr|rrr|rrr}
\tablecaption{Model uv-visibility Fitting Statistics}
\tablehead{
\colhead{Fractional Alignment} & \multicolumn{3}{c}{$<$100~au} & \multicolumn{3}{c}{$<$10,000~au} & \multicolumn{3}{c}{$>$500~au}\\
\colhead{} & \colhead{\%$_{KS}$} & \colhead{\%$_{AD}$} & \colhead{\%$_{ES}$} & \colhead{\%$_{KS}$} & \colhead{\%$_{AD}$} & \colhead{\%$_{ES}$} & \colhead{\%$_{KS}$} & \colhead{\%$_{AD}$} & \colhead{\%$_{ES}$}
}
\startdata
C$_{1.00}$ UC$_{0.00}$ & 100.0 & 100.0 & 32.5 & 0.0 & 0.0 & 0.7 & 0.0 & 0.0 & 19.9\\
C$_{0.90}$ UC$_{0.10}$ & 100.0 & 100.0 & 2.1 & 0.9 & 0.1 & 48.3 & 46.6 & 5.3 & 99.9\\
C$_{0.80}$ UC$_{0.20}$ & 100.0 & 100.0 & 0.0 & 85.6 & 80.8 & 84.1 & 100.0 & 100.0 & 100.0\\
C$_{0.70}$ UC$_{0.30}$ & 100.0 & 100.0 & 0.0 & 100.0 & 100.0 & 99.9 & 100.0 & 100.0 & 100.0\\
C$_{0.60}$ UC$_{0.40}$ & 99.5 & 99.0 & 0.0 & 99.9 & 100.0 & 99.2 & 100.0 & 100.0 & 99.4\\
C$_{0.50}$ UC$_{0.50}$ & 69.7 & 82.6 & 0.0 & 91.6 & 99.0 & 93.4 & 98.0 & 100.0 & 94.1\\
C$_{0.40}$ UC$_{0.60}$ & 48.1 & 44.2 & 0.0 & 51.3 & 72.4 & 73.8 & 55.2 & 90.5 & 68.6\\
C$_{0.30}$ UC$_{0.70}$ & 33.4 & 5.7 & 0.0 & 7.9 & 8.2 & 37.6 & 9.0 & 3.8 & 26.8\\
C$_{0.20}$ UC$_{0.80}$ & 6.9 & 0.5 & 0.0 & 1.0 & 0.1 & 13.2 & 0.5 & 0.0 & 7.7\\
C$_{0.10}$ UC$_{0.90}$ & 3.4 & 0.0 & 0.0 & 0.1 & 0.0 & 4.4 & 0.0 & 0.0 & 2.5\\
C$_{0.00}$ UC$_{1.00}$ & 2.7 & 0.0 & 0.0 & 0.0 & 0.0 & 2.2 & 0.0 & 0.0 & 1.3\\
\enddata
\tablecomments{The fractional alignment column details the fractional ratio of aligned vs randomly aligned as indicated by the subscripts. \%$_{KS}$, \%$_{AD}$, and \%$_{ES}$ correspond to the \% of resampled observation dot products that cannot reject the null hypothesis in favour of the alternative for Kolmogorov-Smirnov, Anderson-Darling, and Epps-Singleton statistical tests respectively. The rejection criteria is evaluated at 0.3\%. The total number of resampled observation dot products are 10,000. The null hypothesis tested is the empirical distributions and the corresponding constructed fractional distributions are drawn from the same underlying distribution.}\label{table:stats10010000500uv}
\end{deluxetable}
\section{Conclusions} \label{sec:conclusion}

We have presented very high resolution observations (\ab8~au) towards 12 Perseus protostellar multiple systems, resolving and detecting 32 sources within the field of view of the targeted sources. We observed the dust continuum at 1.3~mm and provide spatial resolution at 3~mm data as well. Our results can be summarized as follows:

\begin{enumerate}
    \item We detected and confirmed all previously detected multiple systems from the VANDAM observations \citep{2018tobinlooneyli, 2016tobinlooney}, except one, and we detected a total of 32 sources. Per-emb-5 is the only source that we reclassified as a single source. 
    \item We detected circum-multiple continuum emission in seven of the 12 systems consistent with other observations at lower resolution.
    \item We statistically characterized our full sample of 11 Perseus protostellar multiples (N$_{pairs}$=21), with separations $<$10,000~au, to be consistent with forming from a combination of gravitational instability and turbulent fragmentation pathways, with the underlying distribution described as 40\%-80\%\space preferentially aligned systems (or conversely 60\%-20\%\space randomly aligned systems).
    \item If we select a compact subsample of the sources, with separations $<$100~au (N$_{pairs}$=7), we found the underlying distribution of at least 40\%\space preferentially aligned and we ruled out distributions that are randomly aligned dominated.
    \item Futhermore, if we select an extended subsample of the sources, with separations $>$500~au (N$_{pairs}$=15), we found the underlying distributions of 40\%-80\%\space preferentially aligned companions are equally likely.
    \item Combining our statistical approach with prior observations, we determined 7 of 12 of our systems are more consistent with disk fragmentation; while 3 systems (and 3 wide companions) are more consistent with turbulent fragmentation. One system, NGC1333 IRAS2B has an ambiguous formation pathway. \added{Determining the formation mechanism via the statistical approach gives the likelihood of the sample being consistent with some underlying distribution of aligned and misaligned disk; whereas, combining this approach with multi-band observations detailing larger scale structures and molecular line features will give the most holistic determination.}

    \item Towards Per-emb-2, we detected the previously reported compact ($a=$24~au) binary Per-emb-2-A/-B, 2 additional possible companions Per-emb-2-C/-D, and additionally a potential fifth companion Per-emb-2-E, embedded within the northern part of the disk.
    \item Per-emb-5 is now resolved and appears as a ringed disk with a single spiral arm extending to the west. The double peaks as appeared in the VLA data were likely associated with bright peaks along the ring and the underlying disk structure was below the surface brightness limit of the VLA.
    \item Toward L1448 IRS3B, we resolved the compact ring of the inner disk and resolved the previously detected IRS3B-A, -B as compact sources just inside and outside of the ring, respectively. Additionally, toward L1448 IRS3B, while faint, we confidently detected an additional continuum source at the geometric center of the inner disk ring. This might be emission surrounding a more central protostellar source that could dominate the gravitational potential given that it is near the two kinematic centers as described with the molecular lines \cso\space and \ceo.
    \item Towards L1448 IRS3A we resolved the disk to be a ring and an unresolved compact continuum source surrounding the central protostar at the geometric center of the ring.
\end{enumerate}

These results, while unable to determine with certainty the relative contribution of the different formation pathways due to the low number of systems, suggest a path for characterizing multiple star formation with future surveys.  To best utilize the methods described, the highest resolution surveys towards larger samples of multiples are required to confidently resolve the most compact scales of the circumstellar disks around the protostars.

\begin{acknowledgments}

NKR, JJT, and NAK gratefully acknowledge support from NSF AST-1814762. LWL acknowledges support from NSF AST-2108794. \added{ZYL is supported in part by NASA 80NSSC20K0533 and NSF AST-2307199.} DMSC is supported by an NSF Astronomy and Astrophysics Postdoctoral Fellowship under award AST-2102405. DMSC is grateful for support from the Max Planck Society. This paper makes use of the following ALMA data: ADS/JAO.ALMA\#2019.1.01425.S and 2017.1.00337.S. ALMA is a partnership of ESO (representing its member states), NSF (USA) and NINS (Japan), together with NRC (Canada), MOST and ASIAA (Taiwan), and KASI (Republic of Korea), in cooperation with the Republic of Chile. The Joint ALMA Observatory is operated by ESO, AUI/NRAO and NAOJ. The National Radio Astronomy Observatory is a facility of the National Science Foundation operated under cooperative agreement by Associated Universities, Inc.

\end{acknowledgments}

\bibliography{ms}{}
\bibliographystyle{aasjournal}

\appendix

\section{Notes on Specific Sources}\label{sec:specificsources}

\subsection{Class 0}

\subsection*{Per-emb-2}\label{sec:per2_app}
We continue our discussion of Per-emb-2, a previously reported close multiple (a$<$50~au). We resolved the compact binary (a\ab80~mas$\approx24$~au) in both the 1.3~mm and 3~mm observations and from our observations further revealed a possible additional 3 compact sources.

\citet{2018tobinlooneyli}\space observed a velocity gradient in \tco\space and \ceo\space centered on the compact binary of Per-emb-2-AB.  Several other studies have also detected the outflow from Per-emb-2-AB \citep[e.g., ][]{2015tobindunhamlooney,2015yenkoch, 2018stephensdunham}, but none have had the resolution or sensitivity to detect which of the two compact components is driving the outflow. Also, there do not appear to be detectable outflows originating from any of the companion sources. 

\subsection*{Per-emb-5}\label{sec:per5_app}
Previous Submillimeter Array (SMA) observations towards Per-emb-5 at 1.3~mm targeting \ceo~(2-1)\space and CO~(2-1)\space molecular lines indicate a rich gas presence in the disk and envelope \citep{2018stephensdunham, 2022heimsoth}. The source also exhibits bi-polar outflows as evident in the SMA data, which are orthogonal to the major axis of the continuum disk in these observations. The \ceo\space observations shown in \citet{2022heimsoth} exhibit a velocity gradient that is in the same direction as the outflow with $\sim$3\arcsec~(900~au) resolution, along the minor axis of the disk that we detect. Thus, the envelope may be influenced by the outflow \citep{2006arcesargent}, or there is infall from a flattened envelope \citep{2011yentakakuwa}. Higher resolution kinematic observations are required to determine at what scale the disk rotation is detectable.

\subsection*{L1448~IRS3B}\label{sec:irs3b_app}
\citet{2021reynoldstran} observed \cso\space in the disk of IRS3B. \cso\ appears to trace well-ordered Keplerian rotation on the scale of the continuum disk \citep[see Figure~4;][]{2021reynoldstran}. Considering only the high velocity Doppler-shifted channels of the \cso\space emission, which should trace scales closest to the center of mass,  the kinematic center of \cso\space coincides with the position of L1448 IRS3B-D. In Figure~\ref{fig:l1448irs3b}, we compile our observations at 1.3~mm and observations by \citet{2021reynoldstran}. We plot the kinematic centers of the \cso\space molecular line observations using the position-velocity diagram fitting technique (i.e. ``1'') and the Bayesian analysis of kinematic flared disk \textit{uv}-visibility modeling (``2'') with the \textit{pdspy} software \citep{2022pdspy}. We also indicate the position of L1448 IRS3B-A,-B, derived from previous observations, which we now resolved as two sources just inside and outside of the ring, respectively. Finally we draw a visual aid indicating the proposed geometric configuration of the inner disk, which now appears as a ring with a faint, but confidently detected, continuum source located at the geometric center of the ring.

\paragraph{NGC1333~IRAS2A}\label{sec:iras2a}
We resolved the binary of NGC1333~IRAS2A and resolved the brighter compact source NGC1333~IRAS2A-A. We also detected faint extended emission around NGC1333~IRAS2A-A at low surface brightness. The compact faint component NGC1333~IRAS2A-B does not appear to have much extended emission and is marginally resolved in these observations. The sources appeared oriented nearly orthogonal to each other in continuum. Bipolar outflows have been observed toward IRAS2A in several studies \citep{2005jorgensenbourke,2013plunkett,2014codellamaury, 2015tobindunhamlooney, 2022jorgensenkuruwita}. The two outflows are nearly orthogonal to each other and appear to originate from the two components of the system. The outflow position angles are approximately orthogonal to the major axis of each presumed driving source. Other molecular gas emission has been characterized toward IRAS2A \citep{2018tobinlooneyli}. There is a possible velocity gradient in \ceo, but the kinematics do not appear highly organized on scales $>$0\farcs3.

\paragraph{Per-emb-17}\label{sec:per17}
We resolved the compact disks toward each component of the binary (a\ab86~au) and detected the faint circumbinary material; circumbinary gas emission was also previously detected by \citet{2018tobinlooneyli}. The northern source, Per-emb-17-A, is brighter and less extended than the southern source, Per-emb-17-B, which appears appears more edge-on. The circumbinary gas detected in \ceo\space and \tco\space show a clear velocity gradient in the same plane as the companions. A bipolar outflow is also seen by \citep{2018tobinlooneyli}, but the extended emission of the outflow is better recovered by observations from \citet{2018stephensdunham}. The outflow position angle is orthogonal to the orientation of the binary system and the orientation of Per-emb-17-A.

\begin{figure}
    \centering
    \includegraphics{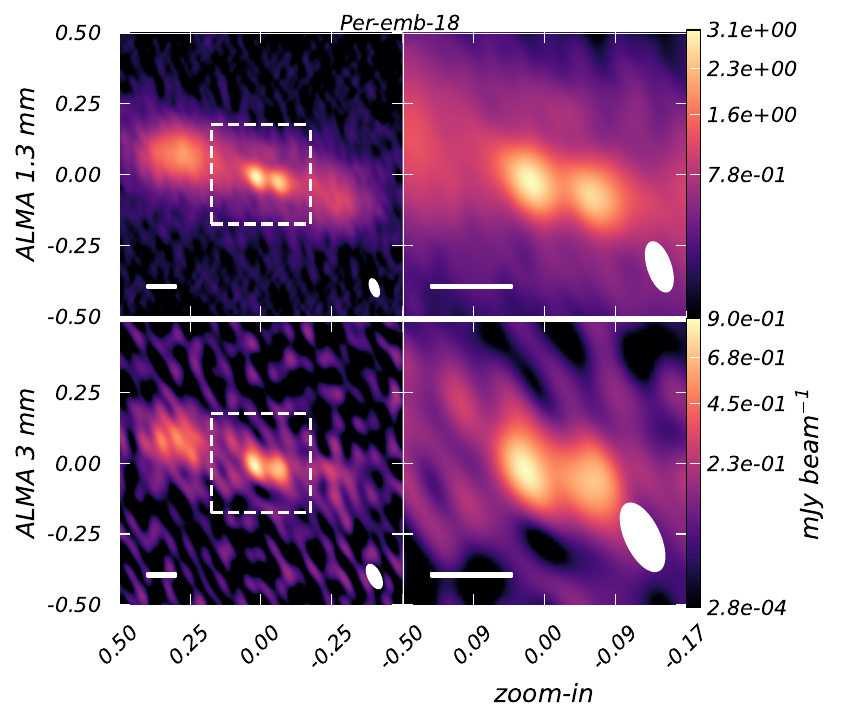}
    \caption{Left side is an 1\farcs0\space image of the Per-emb-18 system, and the right column is a \ab4$\times$~zoom-in. The upper panels are Briggs weighted with robust parameter 1, clean images from the 1.3~mm ALMA data. The bottom panels are super-uniform clean images from the 3~mm ALMA data.  We resolve the compact inner binary of Per-emb-18 and resolve the circumbinary disk. We report in both 1.3~mm and 3~mm observations the circumbinary disk is asymmetric in flux, with the east side being enhanced as compared to the west side. The compact inner binary is located at the geometric center of the circumbinary disk. The compact disks of the individual sources remain unresolved in both observations. A 0\farcs1~(30~au)~scalebar is shown in the lower left and the representative beam is in the lower right. The colormap is square-root scaled.}
    \label{fig:per18}
\end{figure}

\paragraph{Per-emb-18}\label{sec:per18}
We detected both the extended circumbinary structure reported in \citet{2018tobinlooneyli} and resolved the separation between the compact binary (a\ab32~au) as reported in \citet{2016tobinlooney}. The binaries are situated in the near geometric center of the circumbinary disk (Figure~\ref{fig:per18}). There is an apparent surface brightness asymmetry that could appear as an azimuthal asymmetry if viewed more face-on \citep[e.g.,][]{2015marel}. When viewed at 9~mm the circumbinary disk appeared one-sided, which could be due to a combination of lower surface brightness sensitivity from the VLA and dust trapping due to a vortex created by the inner binary pair \citep{2015marel}. The binaries themselves are similar in physical structure and brightness between the 1.3~mm and the 3~mm observations. The circumbinary material is shown to have a clear velocity gradient in \tco, \ceo, and \htco\space in \citet{2018tobinlooneyli}. These velocity gradients are orthogonal to the outflow traced by \co\ \citep{2018stephensdunham, 2018tobinlooneyli, 2022heimsoth}.

\paragraph{Per-emb-21}\label{sec:per21}
Within the same pointing of Per-emb-18, Per-emb-21 is detected 14\farcs~(4230~au) from Per-emb-18. The disk surrounding Per-emb-21 is marginally resolved.
In lower resolution observations, there appears to be \ceo\space emission connecting Per-emb-18 to Per-emb-21, and in \citet{2022heimsoth} they found that there is a $\sim$1~\kms\ line of sight velocity difference between the two sources. Rotation on $<$100~au scales has not yet been detected toward Per-emb-21.

Per-emb-18 and Per-emb-21 also appear to have entangled CO outflows as reported in \citet{2018stephensdunham}. The Per-emb-21 outflow appears much brighter and wider than the Per-emb-18 outflow. The wide-angle (\ab90\deg) outflow spans \ab9000~au and encapsulates both sources in the plane of the sky. However, our reported position angles of the disk major axis are consistent with being orthogonal to the CO outflow emission as reported in \citet{2018stephensdunham}.

\paragraph{Per-emb-22}\label{sec:per22}
Towards Per-emb-22 we marginally resolved the compact disks around each compact (a\ab263~au) source, and we detected a connecting ``bridge'' between the sources in continuum. Additionally, Per-emb-22-B, appears to have an asymmetric oblate disk, extending towards the south and has a large, possibly rotating circumbinary structure in continuum and molecular lines \citep{2015tobinlooneywilner}. The velocity gradient was more evident at $\sim$1\arcsec (300~au) resolution than at $\sim$0\farcs3~(90~au) resolution in \citet{2018tobinlooneyli}. There is a clear outflow from Per-emb-22 \citep{2018stephensdunham} that is at a $\sim$45\degr\ angle with respect to the plane of the binaries. Higher resolution maps of CO from \citet{2018tobinlooneyli} seem to show that the more prominent outflow originates from Per-emb-22-B while there might be some evidence for a second outflow from Per-emb-22-A.

\paragraph{L1448~IRS3C}\label{sec:irs3c}
We resolved both disks of the compact (a\ab72~au) binary L1448~IRS3C. We detected faint circum-binary continuum emission at the limit of our observations. The disks appear nearly aligned and the southern source is much brighter than the northern source. This system is also a wide companion to the L1448~IRS3A/IRS3B system, located \ab5300~au away. 

\citet{2018tobinlooneyli}\space reported observations of L1448~IRS3C with disk tracing molecules \ceo\space and \tco, and found position angles of the velocity gradients corresponding to 220\deg, consistent with the major axes of the disk we observed here. There is also a velocity gradient in the circumbinary emission, consistent with orientation of the binary, in both \tco\space and \ceo. Outflows have been observed 
from L1448~IRS3C \citep{2015leedunham, 2015tobinlooneywilner}, but even the highest resolution \co\space maps thus far from \citet{2018tobinlooneyli} only detect a single outflow.

\paragraph{SVS13~A}\label{sec:svs13a}
is a compact (a\ab92~au) binary with prominent spiral arms that were also seen in \citet{2018tobinlooneyli} and \citet{2022diazrodriguezanglada}. We detected the circumbinary one-armed spiral continuum and resolved each compact circumstellar disk. We also detected the unresolved wide (a\ab1830~au) companion SVS13~A2. Additionally, we detected the wide (a\ab4770~au) companion SVS13~B. We resolved SVS13~B, but due to limited sensitivity far out in the primary beam, the image is very noisy. \citet{2018tobinlooneyli} found apparent rotation in the circumbinary spiral structure, while higher resolution
data from \citet{2022diazrodriguezanglada} found rotation across the two binary sources and circumstellar rotation from one.

\citet{2018stephensdunham}\space found the outflow position angle to be 150\deg$\pm10$\deg, which is approximately orthogonal to our reported disk position angle of \ab70\deg. \citet{2018tobinlooneyli}\space found the disk position angle to be \ab220\deg\space(180\deg\space symmetry yields angle \ab50\deg) by evaluating the angle of the velocity gradient in the disk tracing molecules \ceo\space and \tco. This is consistent with the average of the position angles for the two compact sources SVS13A-AB.

\paragraph{RAC1999 VLA20}\label{sec:rac1999}
We detected the continuum source RAC1999~VLA20. However it is believed to be an extra-galactic source \citep[][]{1999rodriguezanglada,2016tobinlooney}\space so the source is not considered for the purpose of the analysis conducted in this paper. A summary of the observed source is provided in Table~\ref{table:rac1999}.

\subsection{Class I/II Protostars}

\paragraph{L1448~IRS1}\label{sec:irs1}
The bright disk of L1448~IRS1-A, the northern source, is well-resolved and is likely the most evolved source in our sample. This source has a bright, \deleted{compact (62$\times$28~mas$\approx$19$\times$8~au) disk and a fainter extended (25~$\times$12~mas$\approx$75$\times$36~au) disk}\added{disk (25~$\times$12~mas$\approx$75$\times$36~au).}. The fainter, southwest companion is marginally resolved orthogonal to the beam. Previous observations of L1448~IRS1 reported a disk position angle of 24\deg\space based on continuum observations \citep{2018tobinlooneyli}, and we report a consistent position angle of 25\deg. Analyzing the CO\space emission, which is tracing the disk toward L1448~IRS1-A, coincides with the  major axis of the disk and is likely tracing only the gas component of the disk and not the outflows.

\paragraph{L1448~IRS3A}\label{sec:irs3a_app}
Within the pointing of L1448~IRS3B, we detected this wide (a\ab2280~au), more evolved companion. Furthermore, we resolved the substructure of the continuum disk to be a clear circumstellar ring. The compact inner disk centered at the geometric center of the ring remains unresolved in these observations, and the relative orientation cannot be determined to be different with respect to the outer ring.

Previous observations of L1448~IRS3A reported the disk position angle to be \ab130\deg\space \citep{2018tobinlooneyli, 2021reynoldstran}\space based on the PA of the ring and apparent orthogonality to the outflows observed by \citet{2015leedunham, 2018tobinlooneyli}. As such, for the purpose of the statistical analysis later in the paper, we assumed the inner compact disk orientation would be comparable to the orientation of the ring. Rotation of the outer disk has been traced in \ceo, \tco, and \cso, and the inner disk, close to the scales of the inner compact continuum source, has been traced in  SO$_2$, consistent with the outer disk \citep{2018tobinlooneyli,2021reynoldstran}. The origin of the ring is beyond the scope of this paper, but we can report that we do not detect a continuum source within the gap between the inner continuum source and the ring.

\paragraph{Per-emb-35}\label{sec:per35}
We marginally resolved the continuum emission toward each source, which are separated by \ab644~au and report no obvious detection of any continuum emission between the two compact sources. 

Observations of the the CO outflows in \citet{2018stephensdunham} show an outflow position angle of 123\deg, and \citet{2018tobinlooneyli} detected what appear to be parallel outflows from each protostar. Due to the compact nature of the circumstellar disks, rotation within them is not clear, though \citet{2018tobinlooneyli} did detect compact blue and red-shifted \htco\space and SO that could be from the disks. With our higher resolution 1.3~mm\space observations, we report a disk position angle of 31\deg\space for Per-emb-35-A, which is consistent with being orthogonal to the outflow as reported in \citet{2018stephensdunham}. 

\paragraph{NGC1333~IRAS2B}\label{sec:iras2b}
The disk around the brighter southern source, NGC1333~IRAS2B-A is well-resolved. The companion source \ab95~au away, is marginally resolved and visually appears mis-aligned to the brighter source. \citet{2018stephensdunham}\space reported the outflow position angle of 24\deg$\pm10$\deg, which is consistent with an earlier map by \citet{2013plunkett}. We reported the disk position angle of IRAS2B-A to be 104\deg, which is nearly orthogonal with the reported outflow position angles. While the companion IRAS2B-B appears relatively aligned ($\theta\approx$30\deg) with the brighter IRAS2B-A source. Clear rotation across the binary or toward either disk has not been observed, although  \citet{2018tobinlooneyli} found emission to generally be concentrated northeast of IRAS2B-A, towards IRAS2B-B, with SO having the greatest likelihood of exhibiting a velocity gradient. \tco\space and H$_{2}$CO\space appear on the north-east side of IRAS2B-A while \ceo\space appears almost entirely on the west side of IRAS2B-B.

\begin{deluxetable}{l|c}
\tablecaption{RAC1999 VLA20}
\tablehead{\multicolumn{2}{c}{\textit{imfit} results}}
\startdata
R.A. & 03:29:4.256 \\
$\delta$ & 31:16:9.03 \\
separation & 9\arcsec~(2700~au) \\
Int. Intensity & 1.80~mJy \\
Peak & 0.48~mJy~beam$^{-1}$ \\
P.A. & 142\deg \\
$\sigma_{major}$ & 73~mas \\
$\sigma_{minor}$ & 31~mas \\
\enddata
\tablecomments{The CASA \textit{imfit} fitting results for the source RAC1999~VLA20, which was detected in the field of SVS13 A. This source is believed to be extra-galactic and not considered part of the SVS13 A system.}
\end{deluxetable}\label{table:rac1999}

\section{Companion Finding Algorithm}\label{sec:companionfinding}
\begin{figure}
    \centering
    \includegraphics[width=4in]{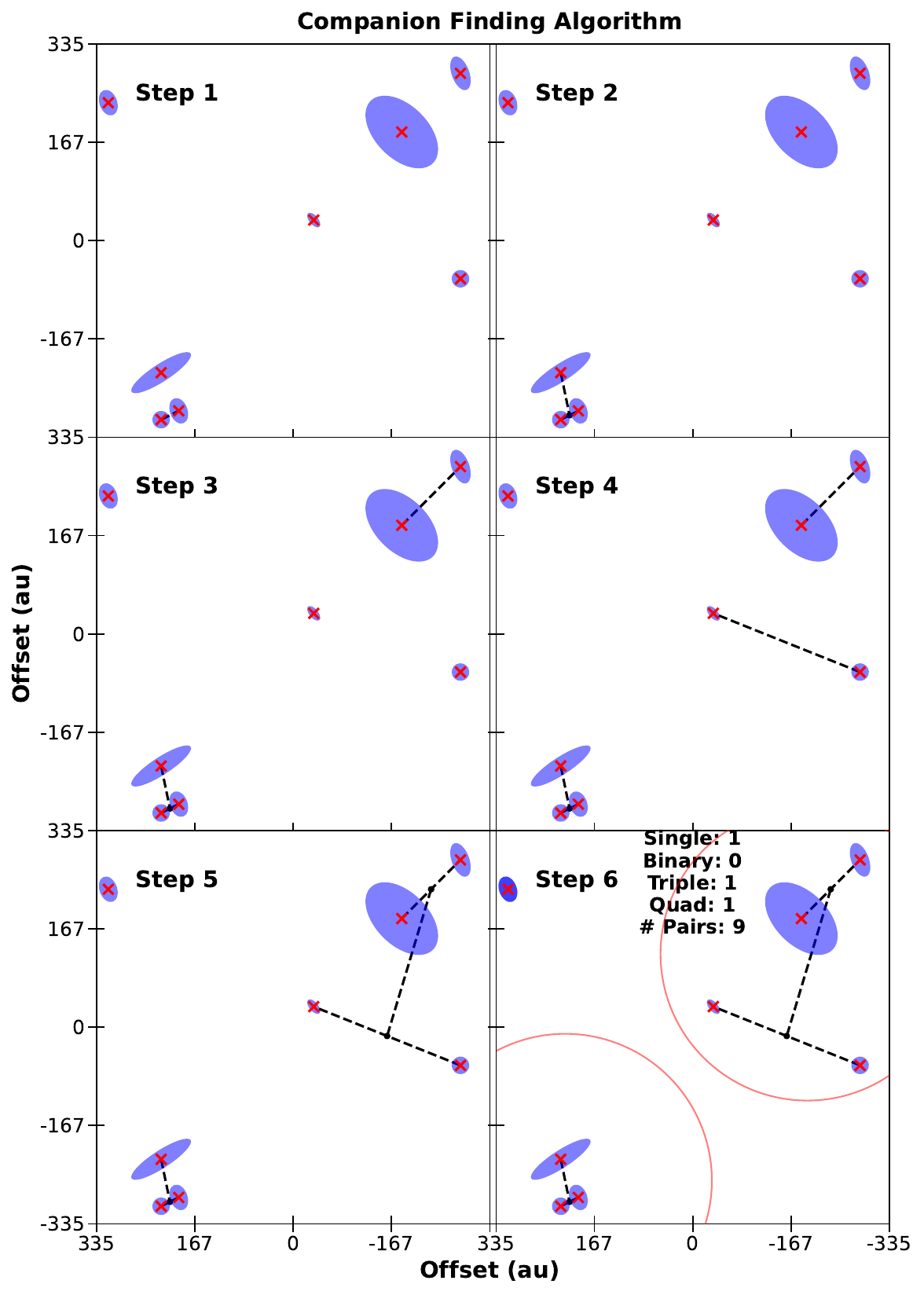}
    \caption{Demonstration of the companion finding algorithm used to associated the observed sources within each system. In this case, the companion maximum distance is 500~au. The algorithm first attempts find the most compact pair within the maximal defined distance and groups them together, re-inserting the geometric center of the group into the source list. The algorithm then finds the most compact pair within the maximum distance allowed and groups these two. In the plot, all of the example sources are shown in red crosses. At ``Step 1'', the algorithm finds the most compact pair. In ``Step 2'' the next most compact pair is a companion to the first binary, thus making a triple source system. The algorithm continues until no more sources meet the distance criteria, arriving at a quad-source system, a triple-source system, and a single source system for this example.}
    \label{fig:convalg}
\end{figure}

\added{The algorithm is given an initial catalog of input positions and derives the L.O.S. separations between each of the sources. }The algorithm first attempts to find the most compact pair within the defined distance criteria and groups them together, re-inserting the geometric center of the group into the source list. The algorithm then re-derives the most compact pair within the maximum distance allowed and groups these two, again re-inserting the new geometric center into the list. In Figure~\ref{fig:convalg}, we show an example case of 8 total sources. indicated as red crosses. At ``Step 1'', the algorithm finds the most compact pair and groups them (System 1). In ``Step 2'' the next most compact pair is a companion to the first binary, thus making a triple source system (System 1). At ``Step 3'', the next most compact pair is a separate pair further away (System 2). At ``Step 4'', another group is made from separate sources (System 3). ``Step 5'', the systems 2 and 3 meet the distance criteria, and thus are merged together to form System 2, now a quad-source system. The algorithm continues until no more sources meet the distance criteria, arriving at a quad-source system, a triple-source system, and a single source system for this particular model. This produces nine pairs of comparison disk orientations.

\section{Gaussian Fitting in the Image Plane}\label{sec:imagefit}
Using the CASA task \textit{imfit}, we fit all of the sources using 2-D Gaussians in the image plane. The initial estimates provided for fitting were single 2-D Gaussians on each individual companion with a Gaussian peak flux density equal to the peak flux density in the image and the beam characteristics as the Gaussian geometry. A few sources needed to be represented with two components, an unresolved point source and an extended 2-D Gaussian: L1448~IRS1, L1448~IRS3B-C, and L1448~IRS3C, NGC1333~IRAS2B. The results of the deconvolved 1.3~mm~image-plane fits are summarized in Table~\ref{table:13mmimfits}, and the results of the deconvolved 3~mm image-plane fits are summarized in Table~\ref{table:3mmimfits}.

The results of Gaussian fitting enable us to analyze the disk geometries of all of the sources. For faint, marginally-resolved ($<$3~beams) or unresolved sources, the recovered structure of the source approaches the inherent beam geometry. For well-resolved sources with bright emission, these effects are less important; however, towards the more compact and marginally-resolved continuum sources, the source parameters derived from Gaussian fits are less-well constrained.

    \begin{longrotatetable}
    \begin{deluxetable}{rlrcccccccccc}
    \setlength\tabcolsep{2.5pt}
    \tablecaption{1.3~mm Image-plane Fitting Results}
    \tablehead{
\colhead{Name} & \colhead{} & \colhead{$\alpha$} & \colhead{$\delta$} & \colhead{Separation} & \colhead{$\sigma_{maj}\times\sigma_{min}$\tablenotemark{a}} & \colhead{$\sigma_{error}$} & \colhead{PA\tablenotemark{a}} & \colhead{PA error} & \colhead{\textit{i}} & \colhead{\textit{i} error} & \colhead{T$_{B}$} & \colhead{Int. Intensity}\\
\colhead{} & \colhead{} & \colhead{(J2000)} & \colhead{(J2000)} & \colhead{(au)} & \colhead{(mas $\times$\space mas)} & \colhead{(mas)} & \colhead{($^{\circ}$)} & \colhead{($^{\circ}$)} & \colhead{($^{\circ}$)} & \colhead{($^{\circ}$)} & \colhead{(K)} & \colhead{(mJy)}
}
    \startdata
L1448 IRS1 & -A & +03:25:09.455 & +30:46:21.84 &  &62 $\times$\space29 &2& 24.7 &1.6 &62.7&4.2 & 80.4 & 9.8\\
\nodata & -B & +03:25:09.418 & +30:46:20.53 & 418 &51 $\times$\space40 &4& 124.0 &13.7 &38.3&6.1 & 30.7 & 4.1\\
Per-emb-2 & -A & +03:32:17.935 & +30:49:47.63 &  &62 $\times$\space28 &45& 19.0 &90.0 &63.4&126.9 & 4.6 & 0.3\\
\nodata & -B & +03:32:17.932 & +30:49:47.70 & 27 &62 $\times$\space28 &45& 19.0 &90.0 &63.4&126.9 & 7.0 & 0.5\\
\nodata & -C & +03:32:17.934 & +30:49:47.29 & 101 &150 $\times$\space116 &18& 121.0 &25.3 &39.4&11.1 & 10.9 & 9.1\\
\nodata & -D & +03:32:17.946 & +30:49:46.30 & 400 &62 $\times$\space28 &45& 19.0 &90.0 &63.4&126.9 & 9.0 & 0.6\\
\nodata & -E & +03:32:17.920 & +30:49:48.19 & 179 &88 $\times$\space65 &27& 157.6 &48.4 &42.6&30.8 & 5.0 & 1.6\\
Per-emb-5 &  & +03:31:20.942 & +30:45:30.19 &  &331 $\times$\space215 &6& 30.4 &2.2 &49.5&2.3 & 57.4 & 176.2\\
NGC1333 IRAS2A & -A & +03:28:55.575 & +31:14:36.92 &  &80 $\times$\space66 &2& 115.6 &7.9 &34.0&2.4 & 264.1 & 83.6\\
\nodata & -B & +03:28:55.568 & +31:14:36.31 & 185 &63 $\times$\space35 &4& 17.0 &4.0 &55.8&7.6 & 66.3 & 10.9\\
Per-emb-17 & -A & +03:27:39.112 & +30:13:02.98 &  &48 $\times$\space37 &1& 117.8 &4.4 &40.6&1.9 & 130.4 & 15.9\\
\nodata & -B & +03:27:39.123 & +30:13:02.73 & 84 &99 $\times$\space52 &3& 158.5 &2.5 &58.6&4.6 & 32.2 & 8.2\\
Per-emb-18+ & -A & +03:29:11.267 & +31:18:30.99 &  &62 $\times$\space36 &10& 63.4 &16.2 &54.3&26.0 & 30.4 & 5.7\\
\nodata & -B & +03:29:11.260 & +31:18:30.97 & 28 &94 $\times$\space40 &12& 77.0 &8.7 &64.8&27.9 & 25.5 & 7.3\\
Per-emb-21+ &  & +03:29:10.674 & +31:18:20.09 &  &50 $\times$\space47 &2& 83.1 &26.1 &21.8&1.8 & 85.4 & 15.9\\
Per-emb-22 & -A & +03:25:22.417 & +30:45:13.16 &  &44 $\times$\space28 &2& 37.0 &4.7 &50.0&5.8 & 130.5 & 12.9\\
\nodata & -B & +03:25:22.359 & +30:45:13.08 & 227 &18 $\times$\space15 &2& 13.7 &52.5 &33.4&8.2 & 74.8 & 4.0\\
L1448 IRS3B+ & -A & +03:25:36.324 & +30:45:14.81 &  &32 $\times$\space12 &8& 27.1 &12.5 &68.6&67.9 & 54.3 & 3.3\\
\nodata & -B & +03:25:36.318 & +30:45:15.06 & 78 &78 $\times$\space55 &4& 34.3 &8.7 &44.6&5.5 & 49.3 & 11.2\\
\nodata & -C & +03:25:36.387 & +30:45:14.64 & 248 &193 $\times$\space168 &4& 44.0 &7.1 &29.2&1.3 & 54.5 & 76.1\\
L1448 IRS3A+ &  & +03:25:36.502 & +30:45:21.83 &  &677 $\times$\space233 &24& 134.2 &1.7 &69.9&7.6 & 4.5 & 30.1\\
L1448 IRS3C & -A & +03:25:35.675 & +30:45:34.02 &  &149 $\times$\space96 &5& 18.9 &3.9 &50.1&4.1 & 115.8 & 73.7\\
\nodata & -B & +03:25:35.678 & +30:45:34.26 & 72 &72 $\times$\space43 &16& 17.9 &37.0 &53.4&30.2 & 79.6 & 13.7\\
Per-emb-35 & -A & +03:28:37.097 & +31:13:30.72 &  &81 $\times$\space33 &2& 31.6 &1.0 &66.0&3.6 & 78.6 & 14.7\\
\nodata & -B & +03:28:37.225 & +31:13:31.67 & 570 &91 $\times$\space27 &2& 48.0 &1.1 &72.8&4.3 & 62.3 & 12.7\\
NGC1333 IRAS2B & -A & +03:28:57.379 & +31:14:15.67 &  &163 $\times$\space97 &2& 104.3 &1.1 &53.7&1.2 & 126.1 & 100.6\\
\nodata & -B & +03:28:57.373 & +31:14:15.98 & 95 &51 $\times$\space42 &6& 82.3 &37.9 &35.1&10.1 & 49.0 & 9.0\\
SVS13A+ & -A & +03:29:03.772 & +31:16:03.71 &  &70 $\times$\space66 &4& 77.3 &47.4 &20.6&2.4 & 199.3 & 57.0\\
\nodata & -B & +03:29:03.748 & +31:16:03.73 & 92 &146 $\times$\space120 &5& 70.6 &9.4 &35.1&2.7 & 102.8 & 85.3\\
SVS13A2+ &  & +03:29:03.391 & +31:16:01.53 &  &49 $\times$\space33 &7& 4.7 &23.1 &48.2&16.1 & 59.0 & 8.6\\
SVS13B+ &  & +03:29:03.085 & +31:15:51.64 &  &218 $\times$\space131 &18& 79.4 &8.5 &53.2&11.2 & 21.4 & 28.0\\
\enddata
    \tablecomments{The Gaussian image-plane fit results utilizing the CASA \textit{imfit} routine. The sources were given estimates to begin the fitting routine but were otherwise not restricted by any bounding values.}
    \tablecomments{The separations are given in units of au, which is derived based on the average distance to the Perseus molecular cloud of 300~pc. Companion separations are defined as the distance from the first target listed.}
    \tablenotetext{a}{The values for PA are defined as the angle of the major axis, oriented east-of-north. The values for $\sigma_{maj}$, $\sigma_{min}$, and PA are deconvolved from the beam.}\label{table:13mmimfits}
    \end{deluxetable}
    \end{longrotatetable}

\begin{longrotatetable}
\begin{deluxetable}{lrrccccccccc}
\tablecaption{3~mm Image-plane Fitting Results}
\tablehead{
\colhead{Name} & \colhead{} & \colhead{$\alpha$} & \colhead{$\delta$} & \colhead{Separation} & \colhead{$\sigma_{maj}\times\sigma_{min}$\tablenotemark{a}} & \colhead{$\sigma_{error}$} & \colhead{PA\tablenotemark{a}} & \colhead{PA error} & \colhead{\textit{i}} & \colhead{T$_{B}$} & \colhead{Int. Intensity}\\
\colhead{} & \colhead{} & \colhead{(J2000)} & \colhead{(J2000)} & \colhead{(mas)} & \colhead{(mas $\times$\space mas)} & \colhead{(mas)} & \colhead{($^{\circ}$)} & \colhead{($^{\circ}$)} & \colhead{($^{\circ}$)} & \colhead{(K)} & \colhead{(mJy)}
}
\startdata
Per-emb-2 & -A & +03:32:17.93 & +30:49:47.73 & & 108 $\times$\space48 & 0& 26.4 & 0.0& 63.3 & 16.2 & 0.7\\
\nodata & -B & +03:32:17.94 & +30:49:47.68 & 85& 108 $\times$\space48 & 0& 26.4 & 0.0& 63.3 & 28.5 & 1.3\\
\nodata & -C & +03:32:17.93 & +30:49:47.34 & 393& 152 $\times$\space82 & 27& 114.2 & 27.7& 57.1 & 17.2 & 3.1\\
\nodata & -D & +03:32:17.92 & +30:49:48.19 & 491& 88 $\times$\space69 & 16& 11.6 & 30.1& 38.0 & 8.7 & 0.4\\
\nodata & -E & +03:32:17.94 & +30:49:46.32 & 1424& 108 $\times$\space48 & 0& 26.4 & 0.0& 63.3 & 5.1 & 0.2\\
Per-emb-5 &  & +03:31:20.94 & +30:45:30.23 & & 264 $\times$\space180 & 10& 28.8 & 4.3& 47.1 & 55.1 & 24.8\\
Per-emb-18 & -A & +03:29:11.26 & +31:18:31.04 & & 34 $\times$\space16 & 18& 97.1 & 82.5& 62.3 & 24.7 & 1.0\\
\nodata & -B & +03:29:11.26 & +31:18:31.02 & 86& 0 $\times$\space0 & 42& 27.0 & 0.0& 0.0 & 20.1 & 0.9\\
Per-emb-21 &  & +03:29:10.67 & +31:18:20.14 & & 57 $\times$\space34 & 6& 166.7 & 11.3& 53.1 & 26.1 & 5.1\\
\enddata
\tablecomments{The Gaussian image-plane fit results utilizing the CASA \textit{imfit} routine. The sources were given estimates to begin the fitting routine but were otherwise not restricted by any bounding values.}
\tablenotetext{a}{The values for PA are defined as the angle of the major axis, oriented east-of-north. The values for $\sigma_{maj}$, $\sigma_{min}$, and PA are deconvolved from the beam.}\label{table:3mmimfits}
\end{deluxetable}
\end{longrotatetable}

\begin{figure}
    \centering
    \includegraphics[width=\textwidth]{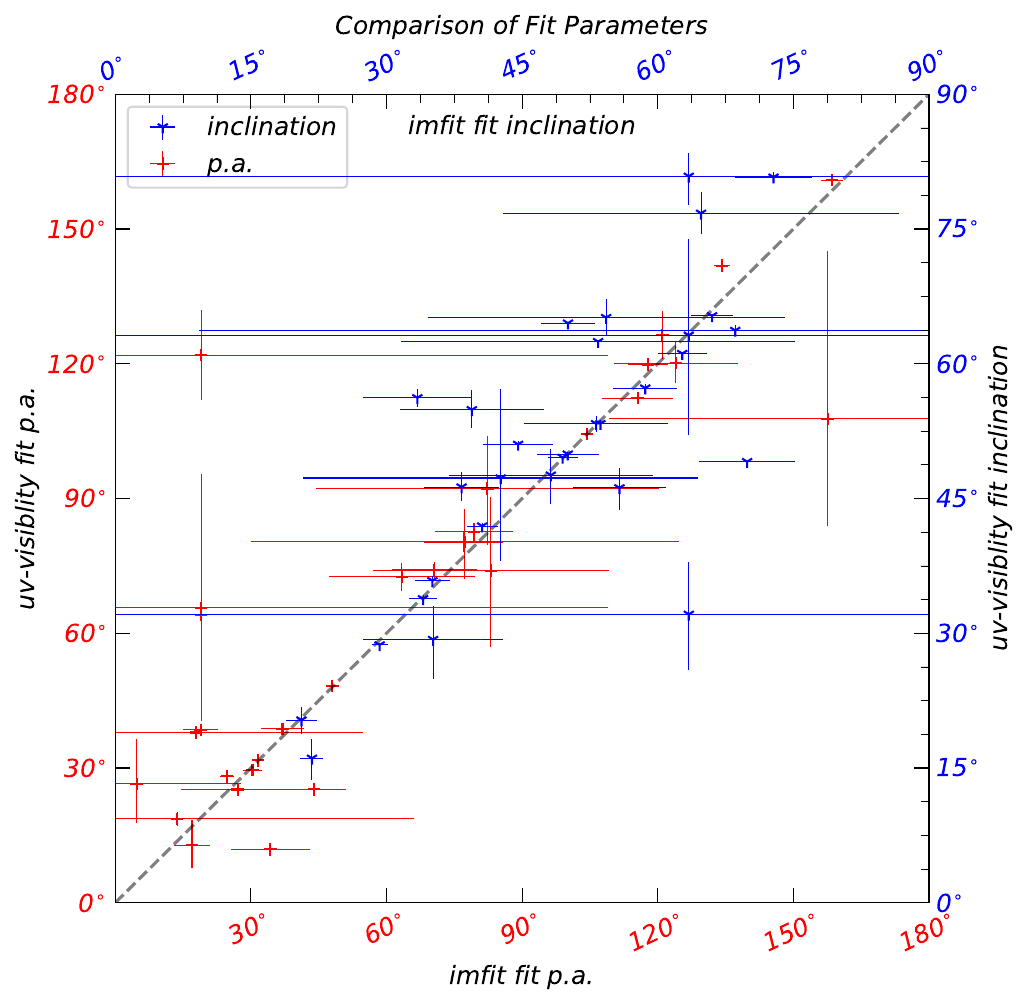}
    \caption{Comparison of the \textit{uv}-visibility fit and imfit derived inclination and position angle orientations for each of the sources. In the cyan, ``Y'' the inclination orientations are shown between 0\deg-90\deg\space and in the red ``$+$'', the position angle orientations are shown between 0\deg-180\deg. The respective errorbars for each fit are also shown, where the \textit{uv}-visibility errors correspond to the isometric 1~$\sigma$\space average uncertainty in the posterior. The one-to-one line demonstrating the two fit methods return the same result is plot in black. There appears to be no clear systematic difference for the two fitting schema\added{, as all sources are within 3~$\sigma$\space of the one-to-one line.}. \deleted{The only significant outlier, Per-emb-2-D, is unresolved in these observations, so the imfit fit would be biased towards the beam geometry.} The \textit{uv}-visibility fit would be able to achieve higher resolution fits as compared to the image-plane fit, thus the \textit{uv}-visibility fit is likely more consistent with the true compact disk orientation.}\label{fig:comparison}
\end{figure}
\begin{figure}
    \centering
    \includegraphics[width=\textwidth]{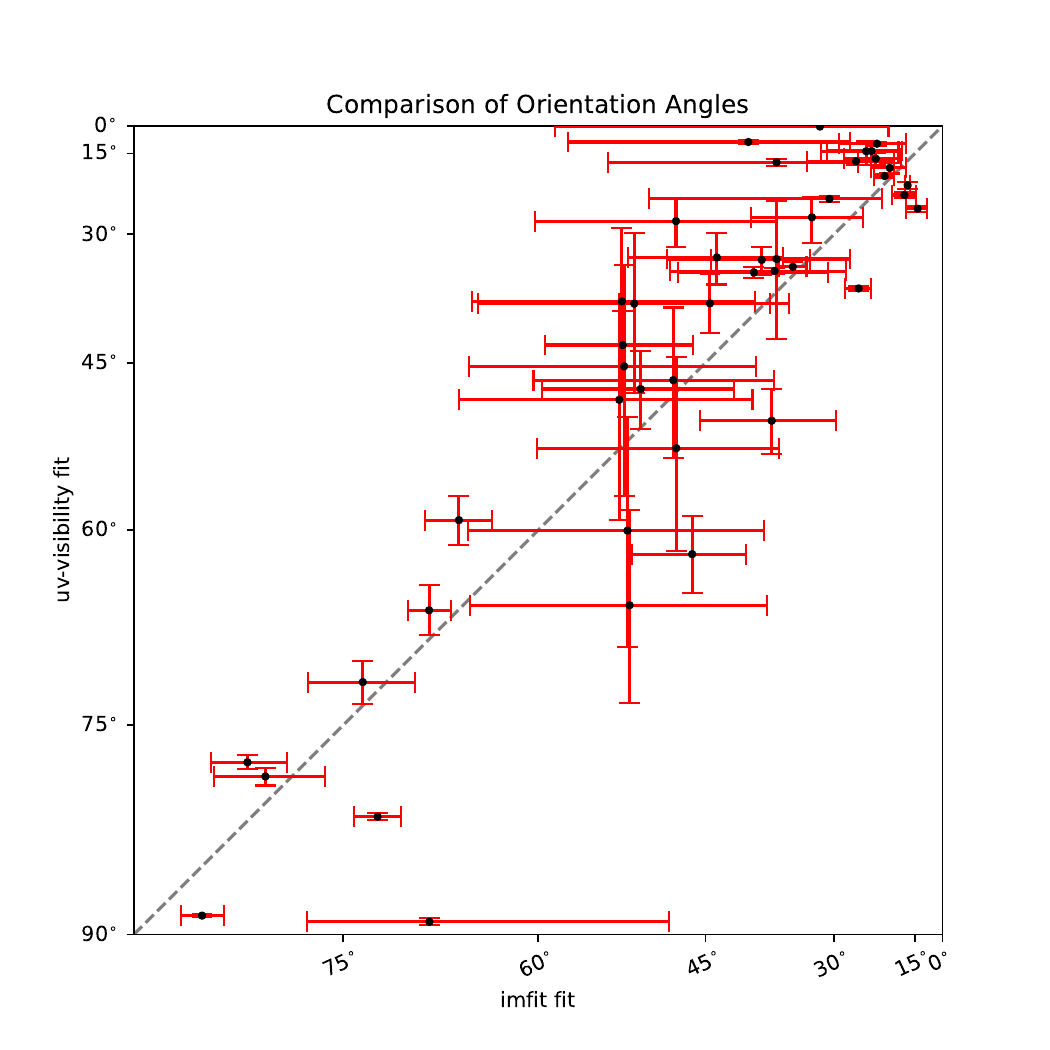}
    \caption{Comparison of the uv-visibility fit and imfit derived dot-product orientations for each of the sources. The red errorbars represent the 1$\sigma$\space confidence intervals while the black points represent the median of the resampled distributions. The one-to-one line demonstrating the two fit methods give comparable results is plot in black, with the upper left half corresponding to targets that appear more aligned with the uv-visibility fitting while the lower right half corresponding to targets that appear more aligned with the image-plane fitting. There appears to be no clear systematic difference for the two fits. The only outlier system L1448 IRS3A.}\label{fig:10000dotprodcomp}

\end{figure}

\section{Comparison of Orientation Vectors}\label{sec:comparisonorientationvectors}
We determine the orientation vectors of each source via independent fitting of the image-plane using the CASA task \textit{imfit} and by fitting 2-D multi-component Gaussians to the \textit{uv}-plane. While the image-plane derived results do not differ significantly from the \textit{uv}-plane results, for the marginally-resolved sources, the image-plane results are less reliable. This is because for marginally-resolved sources, the inherent shape of any recovered source will be nearly identical to the beam geometry, albeit this affect is mitigated in using the deconvolved values, but the deconvolved values are less accurate at low S/N. We show the \textit{uv}-plane and image-plane fit inclination and position angles in Figure~\ref{fig:comparison}. The two techniques are in good agreement within the observational uncertainties.

With the inclination and position angle of each of the companions within a system, we can determine the relative orientations of each of the sources. \added{We compute the relative orientation pairs in a way consistent with the companion finding algorithm previously described by first comparing the most compact binaries within each system. For sources within a multi-body system which are not in a compact configuration, a source from the compact binary is chosen for comparison. This methodology is referred to as ``Method A'' in this work. Similar studies of synthetic simulations compared the orientation of the compact circumstellar disks against the known bound orbit of the constructed protostars \citep{2018bate}. In our observations, for sources with a typical separation of $>50$~au, the orbital time would be $>$300~years, thus the orbital parameters are not well constrained. While other observational studies compared outflow angles to the relative orientation of the larger filamentary structure \citep{2017stephensdunham}, it is known the outflow angles can be misaligned from the disk orientation axis \citep{2016offnerdunham,2022ohashi}.}

\added{One can conceive of several other methods to determine relative orientation parings and to verify our results, we explore these methods.``Method B'' could compute a pairwise permutation (N choose 2 comparisons) of all sources or a ``Method C''  designate a particular protostar as a ``reference'' (perhaps the geometric center or most massive protostar if known) and compute every other source against this primary only once. The result of Method C is highly sensitive to which particular source is determined to be the ``reference'' and any resampling of the chosen binary would result in the same pairs as Method B. We chose Method A for our analysis, but the results presented in this work do not change significantly for any of the methods.}

Given two unit vectors ($\vec{v}_{1}$\space and $\vec{v}_{2}$), defined in spherical coordinates with the angles \textit{i} (defined such that 0\deg\space is face-on) and \textit{p.a.} (defined east-of-north), we can define the magnitude of the relative orientation (the dot product; $\vec{v}_{1}\cdot\vec{v}_{2}$) of these two vectors as $\vec{v}_{1}\cdot\vec{v}_{2}=sin(i_{1})sin(i_{2})cos(p_{1} - p_{2}) + cos(i_{1})cos(i_{2})$. 
Furthermore, with continuum-only detections at this resolution,  we cannot differentiate the position angle of the disk from $+$180\deg. We also cannot differentiate if the disk is ``face-on'' versus ``anti-face-on'' ($+$90\deg), we normalize the inclinations to be 0-90\deg\space and normalize the position angles to be 0-180\deg. 

We perform the same statistical analysis on the image-plane fits and provide the results in Table~\ref{table:stats10010000500im}. We show a comparison of the derived orientations with the \textit{uv}-plane and image-plane fitting techniques in Figure~\ref{fig:10000dotprodcomp}. The points are the median value of the resampled dot products when considering the observation and fitting uncertainties. The only source that stands out is L1448 IRS3A, which has elevated uncertainties due to restricting the \textit{uv}-plane and image-plane fitting to only consider the larger scale structure of the ring.

We summarize our results in Table~\ref{table:imfituvfitcomparison} which details the median separation and the median derived dot-product, considering observation uncertainties, for each of the companions. We also indicate if the particular pair is consistent with being aligned within 30\deg\space as a \checkmark\space or if the system is not aligned as a ``X''.

\begin{deluxetable}{lrrr|rrr|rrr}
\tablecaption{Model image-plane Fitting Statistics}
\tablehead{
\colhead{Fractional Alignment} & \multicolumn{3}{c}{$<$100~au} & \multicolumn{3}{c}{$<$10,000~au} & \multicolumn{3}{c}{$>$500~au}\\
\colhead{} & \colhead{\%$_{KS}$} & \colhead{\%$_{AD}$} & \colhead{\%$_{ES}$} & \colhead{\%$_{KS}$} & \colhead{\%$_{AD}$} & \colhead{\%$_{ES}$} & \colhead{\%$_{KS}$} & \colhead{\%$_{AD}$} & \colhead{\%$_{ES}$}
}
\startdata
C$_{1.00}$ UC$_{0.00}$ & 77.7 & 51.6 & 40.8 & 0.0 & 0.0 & 4.2 & 0.3 & 0.0 & 35.6\\
C$_{0.90}$ UC$_{0.10}$ & 95.8 & 98.6 & 16.8 & 0.9 & 0.1 & 10.6 & 18.2 & 5.8 & 55.0\\
C$_{0.80}$ UC$_{0.20}$ & 99.8 & 100.0 & 6.8 & 29.1 & 21.4 & 28.7 & 78.2 & 70.6 & 67.3\\
C$_{0.70}$ UC$_{0.30}$ & 100.0 & 100.0 & 5.0 & 95.6 & 97.0 & 73.6 & 99.9 & 100.0 & 87.6\\
C$_{0.60}$ UC$_{0.40}$ & 99.9 & 99.9 & 4.3 & 100.0 & 100.0 & 87.5 & 100.0 & 100.0 & 89.4\\
C$_{0.50}$ UC$_{0.50}$ & 98.1 & 99.5 & 3.8 & 98.5 & 99.6 & 86.5 & 98.6 & 99.7 & 84.9\\
C$_{0.40}$ UC$_{0.60}$ & 90.5 & 97.5 & 3.5 & 90.2 & 94.1 & 79.8 & 88.4 & 94.1 & 77.4\\
C$_{0.30}$ UC$_{0.70}$ & 80.4 & 86.9 & 3.3 & 67.6 & 67.3 & 67.5 & 64.0 & 62.7 & 64.2\\
C$_{0.20}$ UC$_{0.80}$ & 70.0 & 74.1 & 3.1 & 45.9 & 38.8 & 52.7 & 45.4 & 37.3 & 50.1\\
C$_{0.10}$ UC$_{0.90}$ & 62.1 & 62.6 & 2.9 & 32.6 & 21.2 & 41.3 & 33.9 & 21.6 & 41.0\\
C$_{0.00}$ UC$_{1.00}$ & 58.5 & 58.5 & 2.9 & 27.6 & 16.5 & 37.2 & 29.8 & 17.5 & 37.3\\
\enddata
\tablecomments{Same as Table~\ref{table:stats10010000500uv}, except using the image-plane results from the deconvolved \textit{imfit}\space task from CASA.}\label{table:stats10010000500im}
\end{deluxetable}
\startlongtable
\begin{deluxetable}{rlccccc}
\tablecaption{Comparison of Derived Dot-products}
\tablehead{
\colhead{Source A} & \colhead{Source B} & \multicolumn{2}{c}{Separation (au)} & \multicolumn{2}{c}{Dot-product} & \colhead{Aligned ($<$30\deg)}\\
\colhead{} & \colhead{} & \colhead{\textit{imfit}} & \colhead{\textit{uv}-plane} & \colhead{\textit{imfit}} & \colhead{\textit{uv}-plane} & \colhead{}
}
\startdata
L1448IRS1-A & L1448IRS1-B & 418$^{0.08}_{0.08}$ & 416$^{0.14}_{0.14}$ & 0.28$^{0.07}_{0.07}$ & 0.31$^{0.03}_{0.03}$ & X\\
L1448IRS3A & L1448IRS3B-A & 2215$^{3.12}_{3.12}$ & 2213$^{0.02}_{0.02}$ & 0.16$^{0.07}_{0.06}$ & 0.2$^{0.01}_{0.01}$ & X\\
 & L1448IRS3B-B & 2151$^{3.06}_{3.06}$ & 2149$^{0.02}_{0.02}$ & 0.14$^{0.05}_{0.05}$ & 0.21$^{0.01}_{0.01}$ & X\\
 & L1448IRS3B-C & 2201$^{2.98}_{2.98}$ & 2196$^{0.03}_{0.03}$ & 0.3$^{0.03}_{0.03}$ & 0.15$^{0.00}_{0.00}$ & X\\
 & L1448IRS3C-A & 4858$^{3.04}_{3.04}$ & 4874$^{0.06}_{0.06}$ & 0.08$^{0.03}_{0.03}$ & 0.02$^{0.00}_{0.00}$ & X\\
 & L1448IRS3C-B & 4904$^{2.18}_{2.18}$ & 4920$^{0.05}_{0.05}$ & 0.37$^{0.30}_{0.15}$ & 0.02$^{0.00}_{0.00}$ & X\\
L1448IRS3B-A & L1448IRS3B-B & 78$^{0.08}_{0.08}$ & 78$^{0.01}_{0.01}$ & 0.89$^{0.05}_{0.06}$ & 0.96$^{0.00}_{0.00}$ & \checkmark\\
 & L1448IRS3B-C & 248$^{0.34}_{0.34}$ & 247$^{0.02}_{0.02}$ & 0.77$^{0.09}_{0.09}$ & 0.82$^{0.01}_{0.01}$ & X\\
 & L1448IRS3C-A & 6285$^{0.36}_{0.36}$ & 6286$^{0.07}_{0.07}$ & 0.92$^{0.04}_{0.05}$ & 0.98$^{0.00}_{0.00}$ & \checkmark\\
 & L1448IRS3C-B & 6346$^{1.45}_{1.45}$ & 6347$^{0.06}_{0.06}$ & 0.76$^{0.13}_{0.22}$ & 0.98$^{0.00}_{0.00}$ & \checkmark\\
L1448IRS3B-B & L1448IRS3B-C & 292$^{0.12}_{0.12}$ & 290$^{0.03}_{0.03}$ & 0.95$^{0.01}_{0.02}$ & 0.91$^{0.00}_{0.00}$ & \checkmark\\
 & L1448IRS3C-A & 6208$^{0.29}_{0.29}$ & 6209$^{0.07}_{0.07}$ & 0.97$^{0.01}_{0.01}$ & 0.9$^{0.00}_{0.00}$ & \checkmark\\
 & L1448IRS3C-B & 6268$^{1.38}_{1.38}$ & 6269$^{0.05}_{0.05}$ & 0.86$^{0.07}_{0.22}$ & 0.91$^{0.00}_{0.00}$ & \checkmark\\
L1448IRS3B-C & L1448IRS3C-A & 6431$^{0.25}_{0.25}$ & 6430$^{0.02}_{0.02}$ & 0.9$^{0.02}_{0.02}$ & 0.8$^{0.00}_{0.00}$ & X\\
 & L1448IRS3C-B & 6490$^{1.33}_{1.33}$ & 6490$^{0.01}_{0.01}$ & 0.79$^{0.09}_{0.13}$ & 0.82$^{0.00}_{0.00}$ & X\\
L1448IRS3C-A & L1448IRS3C-B & 72$^{1.15}_{1.15}$ & 72$^{0.01}_{0.01}$ & 0.85$^{0.08}_{0.33}$ & 1.0$^{0.00}_{0.00}$ & \checkmark\\
NGC1333IRAS2A-A & NGC1333IRAS2A-B & 185$^{0.26}_{0.26}$ & 186$^{0.25}_{0.25}$ & 0.4$^{0.04}_{0.04}$ & 0.51$^{0.03}_{0.03}$ & X\\
 & NGC1333IRAS2B-A & 9424$^{0.07}_{0.07}$ & 9424$^{0.02}_{0.02}$ & 0.93$^{0.01}_{0.01}$ & 0.94$^{0.00}_{0.00}$ & \checkmark\\
 & NGC1333IRAS2B-B & 9346$^{0.19}_{0.19}$ & 9346$^{0.19}_{0.19}$ & 0.91$^{0.04}_{0.06}$ & 0.97$^{0.01}_{0.02}$ & \checkmark\\
NGC1333IRAS2A-B & NGC1333IRAS2B-A & 9320$^{0.18}_{0.18}$ & 9320$^{0.22}_{0.22}$ & 0.37$^{0.03}_{0.03}$ & 0.4$^{0.03}_{0.03}$ & X\\
 & NGC1333IRAS2B-B & 9243$^{0.08}_{0.08}$ & 9243$^{0.06}_{0.06}$ & 0.63$^{0.11}_{0.12}$ & 0.67$^{0.05}_{0.05}$ & X\\
NGC1333IRAS2B-A & NGC1333IRAS2B-B & 95$^{0.23}_{0.23}$ & 95$^{0.16}_{0.16}$ & 0.84$^{0.06}_{0.08}$ & 0.89$^{0.03}_{0.03}$ & X\\
Per-emb-17-A & Per-emb-17-B & 84$^{0.35}_{0.35}$ & 83$^{0.08}_{0.08}$ & 0.82$^{0.02}_{0.02}$ & 0.83$^{0.01}_{0.01}$ & X\\
Per-emb-18-A & Per-emb-18-B & 28$^{0.45}_{0.45}$ & 25$^{0.09}_{0.09}$ & 0.91$^{0.04}_{0.05}$ & 0.97$^{0.01}_{0.01}$ & \checkmark\\
 & Per-emb-21 & 3987$^{0.51}_{0.51}$ & 3985$^{0.12}_{0.12}$ & 0.79$^{0.08}_{0.09}$ & 0.64$^{0.04}_{0.04}$ & X\\
Per-emb-18-B & Per-emb-21 & 3966$^{0.71}_{0.71}$ & 3966$^{0.17}_{0.17}$ & 0.69$^{0.07}_{0.07}$ & 0.47$^{0.05}_{0.05}$ & X\\
Per-emb-2-A & Per-emb-2-B & 26$^{0.46}_{0.46}$ & 25$^{0.57}_{0.57}$ & 0.62$^{0.17}_{0.19}$ & 0.78$^{0.09}_{0.11}$ & X\\
 & Per-emb-2-C & 126$^{0.57}_{0.57}$ & 129$^{0.38}_{0.38}$ & 0.67$^{0.13}_{0.17}$ & 0.6$^{0.11}_{0.13}$ & X\\
 & Per-emb-2-D & 424$^{0.17}_{0.17}$ & 430$^{0.61}_{0.61}$ & 0.61$^{0.17}_{0.20}$ & 0.5$^{0.14}_{0.14}$ & X\\
 & Per-emb-2-E & 153$^{1.27}_{1.27}$ & 150$^{0.25}_{0.25}$ & 0.61$^{0.16}_{0.19}$ & 0.7$^{0.13}_{0.16}$ & X\\
Per-emb-2-B & Per-emb-2-C & 103$^{0.07}_{0.07}$ & 109$^{0.14}_{0.14}$ & 0.67$^{0.12}_{0.17}$ & 0.69$^{0.09}_{0.10}$ & X\\
 & Per-emb-2-D & 401$^{0.36}_{0.36}$ & 408$^{0.09}_{0.09}$ & 0.61$^{0.17}_{0.20}$ & 0.41$^{0.12}_{0.12}$ & X\\
 & Per-emb-2-E & 178$^{0.78}_{0.78}$ & 174$^{0.79}_{0.79}$ & 0.6$^{0.16}_{0.19}$ & 0.78$^{0.09}_{0.12}$ & X\\
Per-emb-2-C & Per-emb-2-D & 299$^{0.44}_{0.44}$ & 300$^{0.23}_{0.23}$ & 0.67$^{0.12}_{0.17}$ & 0.88$^{0.03}_{0.03}$ & X\\
 & Per-emb-2-E & 276$^{0.69}_{0.69}$ & 278$^{0.64}_{0.64}$ & 0.79$^{0.09}_{0.14}$ & 0.84$^{0.07}_{0.10}$ & X\\
Per-emb-2-D & Per-emb-2-E & 575$^{1.1}_{1.1}$ & 578$^{0.87}_{0.87}$ & 0.6$^{0.16}_{0.20}$ & 0.66$^{0.12}_{0.15}$ & X\\
Per-emb-22-A & Per-emb-22-B & 227$^{0.04}_{0.04}$ & 227$^{0.04}_{0.04}$ & 0.79$^{0.10}_{0.21}$ & 0.96$^{0.00}_{0.00}$ & \checkmark\\
Per-emb-35-A & Per-emb-35-B & 570$^{0.06}_{0.06}$ & 570$^{0.48}_{0.48}$ & 0.96$^{0.00}_{0.00}$ & 0.93$^{0.00}_{0.00}$ & \checkmark\\
SVS13A-A & SVS13A-B & 92$^{0.32}_{0.32}$ & 92$^{0.07}_{0.07}$ & 0.92$^{0.03}_{0.04}$ & 0.96$^{0.01}_{0.01}$ & \checkmark\\
 & SVS13A2 & 1603$^{0.14}_{0.14}$ & 1603$^{0.33}_{0.33}$ & 0.71$^{0.10}_{0.11}$ & 0.78$^{0.04}_{0.04}$ & X\\
 & SVS13B-A & 4483$^{1.91}_{1.91}$ & 4482$^{0.88}_{0.88}$ & 0.78$^{0.06}_{0.06}$ & 0.83$^{0.02}_{0.02}$ & X\\
SVS13A-B & SVS13A2 & 1522$^{0.24}_{0.24}$ & 1522$^{0.27}_{0.27}$ & 0.72$^{0.08}_{0.11}$ & 0.84$^{0.03}_{0.03}$ & X\\
 & SVS13B-A & 4433$^{1.65}_{1.65}$ & 4433$^{0.82}_{0.82}$ & 0.93$^{0.02}_{0.02}$ & 0.95$^{0.01}_{0.01}$ & \checkmark\\
SVS13A2 & SVS13B-A & 3190$^{1.35}_{1.35}$ & 3190$^{0.41}_{0.41}$ & 0.6$^{0.09}_{0.10}$ & 0.73$^{0.04}_{0.04}$ & X\\
\enddata
\tablecomments{The separations are given in units of au, which is derived based on the average distance to the Perseus molecular cloud of 300~pc.}
\tablecomments{Compiled comparison table between the \textit{uv}-plane derived results and the image-plane derived results for each of the sources. The ``source A'' and ``source B'' columns correspond to each pair where the comparison was performed to derive the orientation vector correlations. Where ``source A'' is not listed, the previously listed ``source A'' is to be assumed. All separations are defined with respect to the ``source A''. The dot-product is given in cosine coordinates. The final column, ``Aligned ($<$30\deg)'' compares the derived dot-product values by averaging the two values and if the value corresponds to $<$30\deg, the sources are considered aligned, while values $>$30\deg\space are considered misaligned.}\label{table:imfituvfitcomparison}

\end{deluxetable}

\section{Statistical Tests}\label{sec:stattest}

Across all scientific fields, many statistical tests have been used to test whether two independent empirical samples were drawn from the same underlying distribution. This procedure of testing the the distributions usually involves solving the inverted hypothesis, whether the ``null hypothesis'' can be rejected. The null hypothesis being that the two distributions are drawn from the same underlying distribution against the alternative hypothesis, that the two distribution are not drawn from the same underling distribution. If we reject the null hypothesis at some significance level, then we can confidently say the underlying distributions are not identical at that significance level. However, careful consideration is needed as proving the inverse is not necessarily proving the distributions are drawn from the same underlying population. That is to say if we cannot reject the null hypothesis in favor of the alternative, that simply means the underlying distributions are consistent with being drawn from the same distribution at a particular significance level, and does not state the two distributions themselves are identical.

The most well-known statistical test is the Kolmogorov–Smirnov test (KS test), in the case of k-samples, tests the empirical cumulative distribution functions (ECDFs) of each sample against the null hypothesis that they were drawn from the same distribution (under the assumption one distribution is continuous). The KS test is sensitive to changes in the mass center and shape of the ECDF by computing the maximum difference of the distributions (higher sensitivity towards CDF center). The Anderson-Darling test (AD test) belongs to a family of quadratic rank tests and thus places more weight on the differences in the tails of the distributions. However, both the KS and AD tests assume the samples are drawn from continuous distributions and that the underlying distribution is stochastically larger than the drawn distribution. This is the so-called directional alternative hypothesis.

The Epps-Singleton test (ES test), is an empirical rank test that compares the characteristic functions (CFs) of the distributions, not the distributions themselves as compared to the KS and AD test. The test performs a quadratic form of differences between the CFs. The ES test is more powerful in that it does not assume the directional-alternative hypothesis or the continuous criteria, as it can be applied to discrete distributions. The primary prerequisite for the ES test, however, is the samples are fully independent, both across samples and within the samples \citep{eps1, eps2}, but otherwise is shown to provide a more robust statistical test.

For the purpose of our analysis, we apply all three tests, KS, AD, and ES test (as implemented in \textit{scipy}), with k-samples, against the empirical distributions of the Perseus samples and the suites of mixture models. To satisfy the conditions of the KS and AD tests' directional alternative hypothesis and continuous prerequisite, we construct well-sampled models of 10,000 systems (each system comprised of at least 2 sources). To statistically characterize the observations, we resample all of the observation parameters ($\alpha$, $\delta$, $\sigma_{major}$, $\sigma_{minor}$, \textit{i}, and \textit{p.a.}) with Gaussian errors representative of the uncertainties of the observations 10,000 times. We then compute the statistic and probability for each of the KS, AD, and ES tests for every resample of the observations against every constructed distribution of fractional ratios. We finally compile the number of resampled distributions that cannot reject the null hypothesis at the 0.3\% (3~$\sigma$) significance level and provide these numbers in Table~\ref{table:stats10010000500uv}. This yields six numbers per mixture ratio suite that describes the number of distributions that cannot reject the null hypothesis.

\end{document}